\documentclass[runningheads,a4paper]{llncs}

\usepackage[english]{babel}
\usepackage{amssymb}

\usepackage[%
rm={oldstyle=false,proportional=true},%
sf={oldstyle=false,proportional=true},%
tt={oldstyle=false,proportional=true,variable=true},%
qt=false%
]{cfr-lm}
\usepackage[T1]{fontenc}
\usepackage{lmodern}

\usepackage{graphicx}
\usepackage{subfig}
\usepackage{url}
\usepackage{paralist}

\usepackage{cite}
\usepackage{CJK}
\usepackage{multirow}
\usepackage[T1]{fontenc}
\usepackage{amstext}
\usepackage{amsmath}
\usepackage{cite}

\usepackage{algorithm,algpseudocode}
\usepackage{algorithmicx}

\usepackage[math]{blindtext}

\usepackage{csquotes}

\usepackage{microtype}

\usepackage{url}
\makeatletter
\g@addto@macro{\UrlBreaks}{\UrlOrds}
\makeatother
\usepackage{xcolor}

\usepackage[
bookmarks=false,
breaklinks=true,
colorlinks=true,
linkcolor=black,
citecolor=black,
urlcolor=black,
pdfpagelayout=SinglePage,
pdfstartview=Fit
]{hyperref}
\usepackage[all]{hypcap}

\usepackage{pdfcomment}

\usepackage[capitalise,nameinlink]{cleveref}
\crefname{section}{Sect.}{Sect.}
\Crefname{section}{Section}{Sections}

\usepackage{xspace}
\DeclareFontFamily{U}{MnSymbolC}{}
\DeclareSymbolFont{MnSyC}{U}{MnSymbolC}{m}{n}
\DeclareFontShape{U}{MnSymbolC}{m}{n}{
    <-6>  MnSymbolC5
   <6-7>  MnSymbolC6
   <7-8>  MnSymbolC7
   <8-9>  MnSymbolC8
   <9-10> MnSymbolC9
  <10-12> MnSymbolC10
  <12->   MnSymbolC12%
}{}
\DeclareMathSymbol{\powerset}{\mathord}{MnSyC}{180}

\begin{document}

%Works on MiKTeX only
%hint by http://goemonx.blogspot.de/2012/01/pdflatex-ligaturen-und-copynpaste.html
%also http://tex.stackexchange.com/questions/4397/make-ligatures-in-linux-libertine-copyable-and-searchable
%This allows a copy'n'paste of the text from the paper
\input glyphtounicode.tex
\pdfgentounicode=1

\title{A Graph-based Push Service Platform}
%If Title is too long, use \titlerunning
%\titlerunning{Short Title}

%Single insitute
\author{Huifeng Guo\inst{1}\thanks{The work is done when Huifeng Guo is an intern in Noah's Ark Lab, Huawei.} \and Ruiming Tang\inst{2} \and Yunming Ye\inst{1}\thanks{Corresponding author} \and Zhenguo Li\inst{2} \and Xiuqiang He\inst{2}}
%If there are too many authors, use \authorrunning
\authorrunning{Huifeng Guo et al.}
\institute{
Department of Computer Science, Shenzhen Graduate School, Harbin Institute of Technology, Shenzhen,  China\\
\email{huifengguo@yeah.net yeyunming@hit.edu.cn}\and
Noah's Ark Lab, Huawei, China\\
\email{\{tangruiming,Li.Zhenguo,hexiuqiang\}@huawei.com}
}

%Multiple insitutes
%Currently disabled
%
\iffalse
%Multiple institutes are typeset as follows:
\author{Firstname Lastname\inst{1} \and Firstname Lastname\inst{2} }
%If there are too many authors, use \authorrunning
%\authorrunning{First Author et al.}

\institute{
Insitute 1\\
\email{...}\and
Insitute 2\\
\email{...}
}
\fi
			
\maketitle

\begin{abstract}
It is well known that learning customers' preference and making recommendations to them from today's information-exploded environment is critical and non-trivial in an on-line system. There are two different modes of recommendation systems, namely \emph{pull-mode} and \emph{push-mode}. The majority of the recommendation systems are pull-mode, which recommend items to users only when and after users enter Application Market. While push-mode works more actively to enhance or re-build connection between Application Market and users. As one of the most successful phone manufactures, both the number of users and apps increase dramatically in Huawei Application Store (also named Hispace Store), which has approximately 0.3 billion registered users and 1.2 million apps until 2016 and whose number of users is growing with high-speed. For the needs of real scenario, we establish a Push Service Platform (shortly, \emph{PSP}) to discover the target user group automatically from web-scale user operation log data with an additional small set of labelled apps (usually around 10 apps), in Hispace Store. As presented in this work, PSP includes distributed storage layer, application layer and evaluation layer. In the application layer, we design a practical graph-based algorithm (named \emph{A-PARW}) for user group discovery, which is an approximate version of partially absorbing random walk. Based on I mode of \emph{A-PARW}, the effectiveness of our system is significantly improved, compared to the predecessor to presented system, which uses Personalized Pagerank in its application layer.
\end{abstract}

\keywords{Random walk; Recommendation; Graph mining}

%%%%%%%%%%%%%%%%%%%%%%%%%%%%%%%%%%%%%%%%%%%%%%%%%%%%%%%%%%%%%%%%%%%%%%%%%%%%%%%

%``something in quotes'' using plain tex or use \enquote{the enquote command}.
\section{Introduction}\label{sec:intro}
%%%%%%%%%%%%%%%%%%%%%%%%%%%%%%%%%%%%%%%%%%%%%%%%%%%%%%%%%%%%%%%%%%%%%%%%%%%%%%%
%\blindtext\todo{Refine me}

With the rapid development of internet infrastructure and smart phones, the life of human beings is more and more closely connected to online applications, such as online shopping \cite{li2008online}, online video sharing \cite{covington2016deep} and etc. However, as a difficult yet crucial task to each online application, learning customers' preference and making a recommendation to them from today's information-exploded environment is a big challenge. Therefore, recommendation system plays a particularly critical role to help customers find what they need or prefer, resulting in enhancing users' experience and increasing revenue for online applications at the same time.

There are two different modes of recommendation systems, namely \emph{pull-mode} and \emph{push-mode}. The majority of the recommendation systems are pull-mode, which recommend items to users only when and after users enter Application Market. The pull-mode recommendation is similar to the case of costumers shopping in a supermarket: shop assistants arrange the products in shelves in certain patterns according to customers' shopping habit, so as to implicitly persuade the customers to purchase more products. Acting as salesman, the push-mode recommendation (which is discussed in this paper) pushes the recommended items to the users actively rather than waiting for the users entering Application Market. The difference between pull-mode and push-mode recommendation is that the former one recommends passively while the latter one works actively, so that the push-mode is able to enhance or re-build the connection between products and inactive or semi-active users\footnote{In the context of Application Market, "active users" refers to the users who visits the Application Market frequently, "inactive users" are the ones left Application Market already and "semi-active users" are the users who do not visit the Application Market often and are leaving but not yet.}.

\begin{figure}[ht]
\begin{center}
\subfloat[Push message]{
\label{fig:push-m}
\includegraphics[width=.24\textwidth]{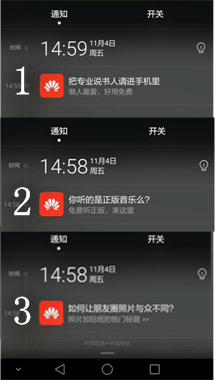}}
\subfloat[Book listening]{
\label{fig:push-1}
\includegraphics[width=.24\textwidth]{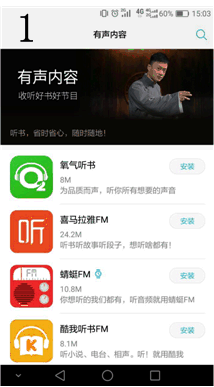}}
\subfloat[Music]{
\label{fig:push-2}
\includegraphics[width=.24\textwidth]{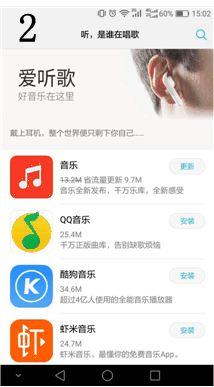}}
\subfloat[Photo editor]{
\label{fig:push-3}
\includegraphics[width=.24\textwidth]{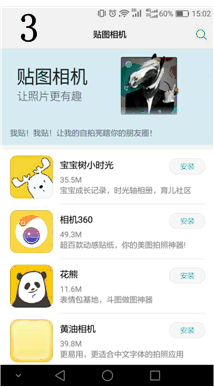}}
\caption{Push Services}
\label{fig:push}
\end{center}
\end{figure}

In Figure~\ref{fig:push}, we present several push activities in Hispace Store \footnote{Hispace Store is the name of Huawei Application Market: \url{http://appstore.huawei.com/hd}. In the rest of this paper, we use Hispace Store to denote Huawei Application Store.}, including book listening, music and photo editor. Through the messages of notification center (presented in Figure~\ref{fig:push-m}), the inactive or semi-active users are able to know some new apps without entering into Hispace Store and download their favorite apps in the display pages after clicking the corresponding notification message (presented in Figure~\ref{fig:push-1}, Figure~\ref{fig:push-2} and Figure~\ref{fig:push-3}).

Compared to the pull-mode recommendation, the push-mode achieves two main advantages, namely (1) reducing the loss of semi-active users and (2) activating the inactivate user, by actively informing the newest updating or interesting messages to the users who do not enter Application Market very often. However, user group selection is a challenging task in push-mode recommendation, since too many unrelated message could disturb users and degrade the user's experience of the phones. In the contrary, accurately targeting users for the corresponding push messages brings huge benefits by upgrading user's experience and increasing the revenue of online applications and Application Market.
%In fact, online applications, such as Google AdWords and Google AdSense, take a huge advantage in accurately targeting the potential user groups for corresponding advertisements[].

However, along with the performance improvement of smart phone and various needs from our daily-life, many apps are created by different developers and installed by different people. Then the number of users and apps increase dramatically these yeas. For example, Hispace Store has around 0.3 billion registered users and 1.2 million apps until 2016. As a result, user group discovery from the web-scale users becomes even more challenging. So we need a platform to support the function of target user group discovery, since the manual labeling is impossible due to missing the ground truth and the large scale of the data.

%Along with the performance improvement of smart phone, people are able to directly install and use various apps on their smart phones. Therefore, the app-user relationship knowledge becomes a new important information source for on-line application recommendation. Moreover, due to various needs from our daily-life, many apps, which are created by different developers, are uploaded to Application Market (such as Google Play, App Store, Hispace Store).
%With the growing capability of smart phone, using various apps on the phone has become a new human daily habit. Therefore, the relationship between the users and apps is forming a new knowledge that can help us to alleviate the ability of on-line advertisement marketing. There is numerous app and user in an app store. the app-user relationship knowledge becomes very important for on-line advertisement marketing.

In response to this real motivation, we have established a Push Service Platform (shortly, \emph{PSP}) for push-mode recommendation, by discovering the target user group. \emph{PSP} includes three layers: distributed storage layer, application layer and evaluation layer. In the application layer, we design a practical graph-based algorithm (named \emph{A-PARW}) for user group discovery, which is a approximate version of \emph{PARW} \cite{wu2012learning}.

In this paper, we present full details of Huawei \emph{PSP}. We firstly overview each layer in Huawei \emph{PSP} generally. We then give a detailed analysis of different user group discovery algorithms theoretically and empirically. We implement the A-PARW algorithm for services with web-scale data size. Finally, we perform our system to several real marketing tasks in Hispace Store, and carried out detailed off-line and on-line test. The results show that, I mode of A-PARW is able to find a target user group effectively. Based on I mode of A-PARW, the effectiveness of our system is significantly improved, compared to PPR (short for personalized pagerank\cite{haveliwala2002topic}). PPR is applied in the predecessor to the current Huawei \emph{PSP}, and the effectiveness of PPR is demonstrated in our previous work \cite{he2015mining}\footnote{xRank, proposed in \cite{he2015mining}, is exactly PPR and is equivalent to D mode of A-PARW.}.

%compared to D mode of A-PARW. D mode of A-PARW is applied in the predecessor to the current Huawei \emph{PSP}, and the effectiveness of D mode is demonstrated in our previous work \cite{he2015mining} (D mode of A-PARW is equivalent to xRank in \cite{he2015mining}).

The rest of the paper is organized as follows. Section~\ref{sec:plat} describes the PSP in general.  Section~\ref{sec:algo} presents the Application Layer of PSP, analyses the differences and similarities of two graph-based algorithms in the application layer in detail and introduces the implementation details of A-PARW in VENUS \cite{cheng2015venus} and PowerGraph \cite{gonzalez2012powergraph}. In Section~\ref{sec:expe} we perform experiments to evaluate the two graph-based algorithms on both public and real-life datasets. We discuss some related works in Section~\ref{sec:relatedwork}. Finally, we conclude the paper in Section~\ref{sec:conclu}.

\section{Platform Overview}\label{sec:plat}

\begin{figure}[ht]
\begin{center}
\includegraphics[width=.96\textwidth]{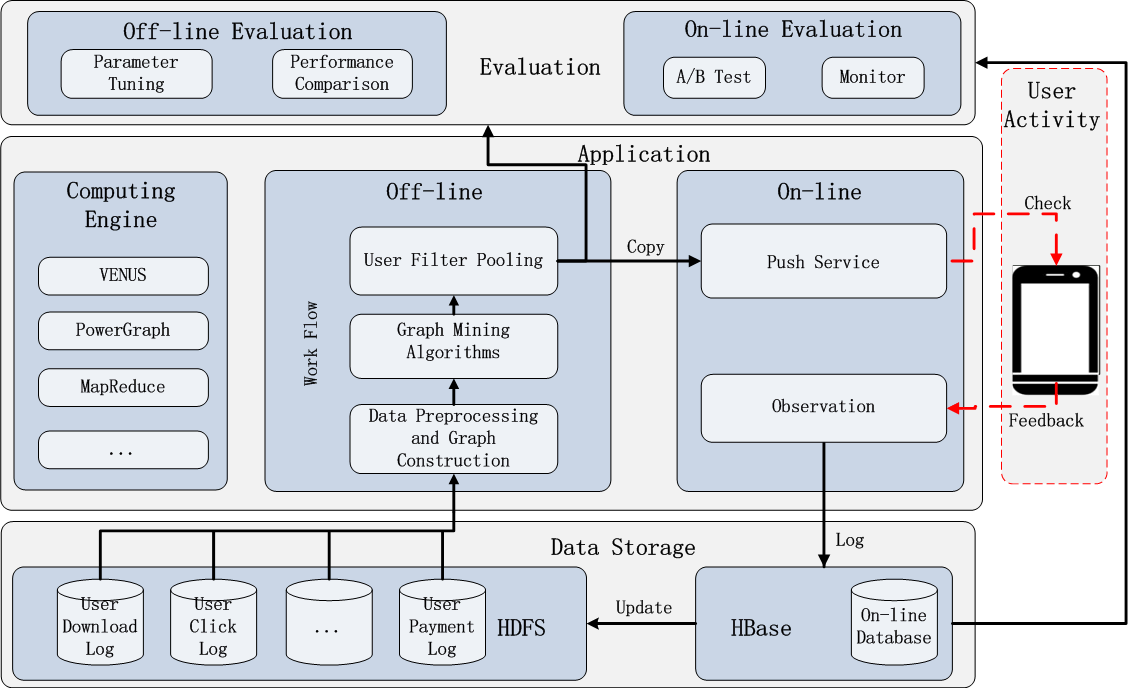}
\caption{PSP Architecture}
\label{fig:platform-overview}
\end{center}
\end{figure}

In order to provide push-mode recommendation to the users, we establish Huawei Push Service Platform (PSP) to select a target user group, to whom send the pre-defined messages. PSP, as presented in Figure \ref{fig:platform-overview}, includes \emph{Distributed Storage Layer}, \emph{Application Layer} and \emph{Evaluation Layer}. Specifically, Storage Layer is built for historical data storage (including the users' download/click/payment historical information) and on-line log caching; Application Layer mines target user group from the graph-structural historical User-APP data in an off-line manner and pushes the pre-defined messages to the selected users' phones; and in Evaluation Layer we tune the parameters, compare the performance of different algorithms both off-line and on-line A/B test. In the rest of this section, we introduce these three layers in more details.

Distributed Storage Layer maintains two database systems for historical data storage and on-line caching. The HDFS (shortening of Hadoop Distributed File System) stores historical data, including users' download, click and payment log data, which is the source data for User-APP bipartite graph construction. While HBase (shortening of Hadoop Database) caches on-line data and users' feedback, which is critical for on-line monitoring and algorithm evaluation. Updating in an incremental way, we use on-line data and users' feedback data in HBase to update the historical data in HDFS periodically. In addition, this layer incorporates a Hadoop cluster to store large-scale datasets and provides parallel data processing.

Application Layer contains the major components (i.e., off-line target users mining and on-line message pushing) of the platform. For the different demands of real project, our Computing Engine of Application Layer supports Graph Engine (such as VENUS and PowerGraph) and distributed computing engines (such as MapReduce). In the off-line part of Application Layer, we first construct User-APP bipartite graph by extracting and preprocessing on the users' historical data stored in HDFS, then mine target user group to send a pre-defined message, and finally we filter the selected users by some beneficial rules. On the other hand, the function of the on-line part is to push the recommendation messages to the selected users by the off-line part and caches users' feedback, that whether he/she clicked (or even downloaded) apps recommended by the sent message, by HBase. Moreover, we will give more details about the Application Layer in Section~\ref{sec:algo} since it is the most challenging part of our system.

This system also includes Evaluation Layer which conducts both off-line and on-line experiments and analyzes the results.  Off-line evaluation compares pre-defined off-line metrics of the results by different algorithms, which helps us to tune the parameters of the algorithms. On-line evaluation monitors and compares the performance of the algorithms which are carefully selected by off-line evaluation. The details of evaluation methods and metrics will be presented in Section~\ref{sec:expe:data-set}.

\section{Application Layer}\label{sec:algo}

\begin{figure}[ht]
\begin{center}
\includegraphics[width=.96\textwidth]{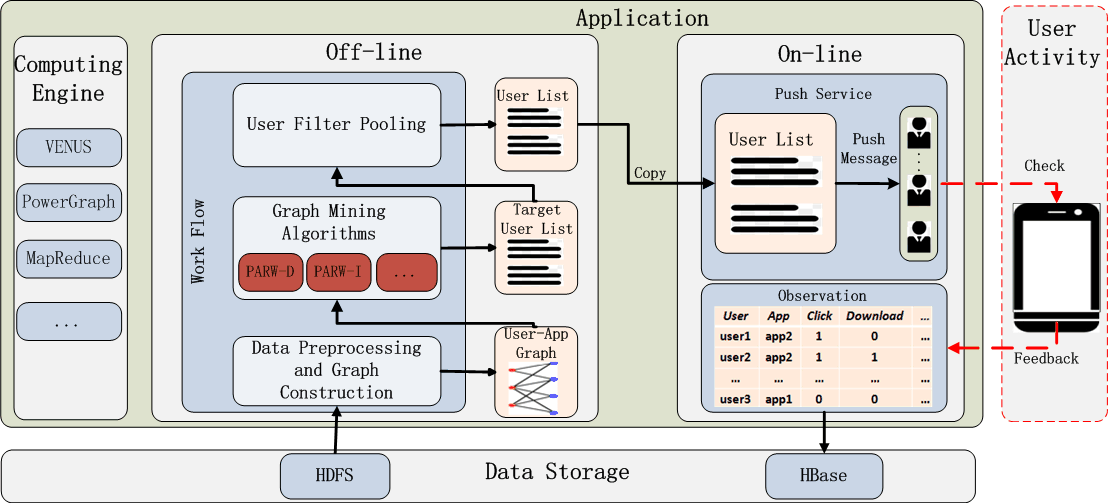}
\caption{Application Layer}
\label{fig:application}
\end{center}
\end{figure}

%\textcolor{red}{To better understand our PSP, in this section, we present the full details of Application Layer (which is the most interesting part) of our PSP. We start with presenting the work flow of Application Layer (shown in Figure~\ref{fig:application}).}
To better understand our PSP, in this section, we present the full details of Application Layer (which is the most interesting part) of our PSP. We start with presenting the work flow of Application Layer (shown in Figure~\ref{fig:application}).

\emph{History Data} $\xrightarrow[Graph \ Construction]{Pre-processing}$ \emph{User-APP Graph} $\xrightarrow{Graph\ Mining}$ \emph{Target User List} $\xrightarrow{User \ Filtering}$ \emph{User List} $\xrightarrow[Observation]{On-line \ Pushing}$ \emph{User Feedback} $\xrightarrow{Logging}$ \emph{Online Log} $\xrightarrow{Updating\ pediodly}$ \emph{History Data}.\\

%\textcolor{red}{The input of PSP is a topic push activity, which is usually a graphic message, describing a group of on-line applications under a certain topic, such as music fans, cook lovers, etc. As an initial step, we provide \emph{Data Preprocessing and Graph Construction} operation (presented in Section~\ref{sec:algo:constr}) to generate User-APP bipartite graph from users' download/click/payment historical data, which is stored in HDFS. Based on this graph, we can mine target user group with the help of \emph{Graph Mining Algorithm} (presented in Section~\ref{sec:algo:graph:algo}). According to some domain knowledge and specific rules, \emph{User Filtering Pooling} filters unneeded users to obtain the final user list (presented in Section~\ref{sec:algo:filter}) and the on-line \emph{Push Service} (presented in Section~\ref{sec:algo:online}) sends the message to the selected users. Then, PSP will cache the users' feedback data and update these data to History data periodically. Moreover, we use diverse computing engines to support a variety of needs in this layer(presented in Section~\ref{sec:algo:engine}).}
The input of PSP is a topic push activity, which is usually a graphic message, describing a group of on-line applications under a certain topic, such as music fans, cook lovers, etc. As an initial step, we provide \emph{Data Preprocessing and Graph Construction} operation (presented in Section~\ref{sec:algo:constr}) to generate User-APP bipartite graph from users' download/click/payment historical data, which is stored in HDFS. Based on this graph, we can mine target user group with the help of \emph{Graph Mining Algorithm} (presented in Section~\ref{sec:algo:graph:algo}). According to some domain knowledge and specific rules, \emph{User Filtering Pooling} filters unneeded users to obtain the final user list (presented in Section~\ref{sec:algo:filter}) and the on-line \emph{Push Service} (presented in Section~\ref{sec:algo:online}) sends the message to the selected users. Then, PSP will cache the users' feedback data and update these data to History data periodically. Moreover, we use diverse computing engines to support a variety of needs in this layer(presented in Section~\ref{sec:algo:engine}).

%\textcolor{red}{In the rest of this section, we introduce the details of Application Layer according to the work flow, which has been briefly described above.}
In the rest of this section, we introduce the details of Application Layer according to the work flow, which has been briefly described above.

\subsection{Data Preprocessing and Graph Construction}\label{sec:algo:constr}

%\textcolor{red}{In this section, we present the details of the first step in the Application Layer, namely data preparation operation.}

In this section, we present the details of the first step in the Application Layer, namely data preparation operation.

\subsubsection{Data Preprocessing}

In the data preprocessing stage, the system can set a series of rules according to demands, such as removing the pre-installed Apps or very popular Apps from raw data before graph construction, because installing such apps are not able to reflect users' interests.

\subsubsection{Graph Construction}

\begin{figure}[ht]
\begin{center}
\includegraphics[width=.33\textwidth]{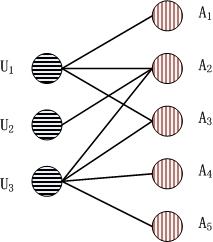}
\caption{An example user-app bipartite graph}
\label{fig:graphexample}
\end{center}
\end{figure}

Based on the data that is being preprocessed, we construct an undirected graph $\mathcal{G}=(\mathcal{U},\mathcal{A},\mathcal{\xi})$, where $\mathcal{U}$ denotes the set of vertices representing \emph{users} and $\mathcal{A}$ is the set of vertices representing $\emph{apps}$, $\mathcal{\xi}$ is a set of edges. Since we only use historical information between users and apps, without the information within users or within apps, the constructed graph $\mathcal{G}$ is a bipartite graph. For instance, in Figure \ref{fig:graphexample}, User vertices are on the left-hand-side and APP vertices are on the right-hand-side. There exists an edge connecting $U_i$ and $A_i$ if user $U_i$ installs app $A_i$. For example, $U_1$ installs three apps $A_1$, $A_2$ and $A_3$, while $A_2$ is installed by all the three users.

We assign uniform IDs to vertices in $\mathcal{A}$ and $\mathcal{U}$. More specifically, vertices in $\mathcal{U}$ are assigned with IDs from 1 to $|\mathcal{U}|$, and vertices in $\mathcal{A}$ are assigned with IDs from $|\mathcal{U}|+1$ to $|\mathcal{U}|+|\mathcal{A}|$. For simplicity, we use $v_i$ to denote vertex with ID $i$. In Figure~\ref{fig:graphexample}, for example, we use $v_1, v_2, v_3$ to denote $U_1, U_2, U_3$ repectively and $v_4, v_5, v_6, v_7, v_8$ to denote $A_1, A_2, A_3, A_4, A_5$ respectively. We denote the adjacency matrix of $\mathcal{G}$ as $\mathcal{W}$, where $\mathcal{W} = [w_{ij}] \in R^{N\times N}$ is a symmetric non-negative matrix of pairwise affinities among vertices (note that $N=|\mathcal{A}|+|\mathcal{U}|$ and $w_{ii}=0$). Let $\mathcal{D}=diag(d_1,d_2,...,d_N)$ with $d_i=\sum_{j}{w_{ij}}$ as the degree of vertex $i$ and define the Laplacian of $\mathcal{G}$ as $\mathcal{L} = \mathcal{D} - \mathcal{W}$. We define the transition matrix of a graph, denoted by $\mathrm{T}$ as:

\begin{eqnarray}\label{OQPSK}
\mathrm{T}(v_i,v_j)=
\begin{cases}
0 &\mbox{if } (v_i,v_j)\not\in \mathcal{\xi}\cr
1/d_i &\mbox{if } (v_i,v_j)\in \mathcal{\xi}
\end{cases}
\end{eqnarray}

For instance, the transition matrix of the graph in Figure \ref{fig:graphexample} is:

\begin{center}
$T=\mathcal{D}^{-1}\cdot \mathcal{W}=\left(
\begin{array} {cccccccc}
0   & 0   & 0   & 1/3 & 1/3 & 1/3 & 0   &0\\
0   & 0   & 0   & 0   & 1   & 0   & 0   &0 \\
0   & 0   & 0   & 0   & 1/4 & 1/4 & 1/4 &1/4\\
1   & 0   & 0   & 0   & 0   & 0   & 0   &0\\
1/3 & 1/3 & 1/3 & 0   & 0   & 0   & 0   &0\\
1/2 & 0   & 1/2 & 0   & 0   & 0   & 0   &0\\
0   & 0   & 1   & 0   & 0   & 0   & 0   &0
\end{array}
\right) $
\end{center}
In some scenarios of push services, the graph has to be updated frequently. Online news recommendation is such a scenario, as the hot spots can be changed at any time. While in some other scenarios, the graph needs to be updated less frequently, such as recommendation in application market. Because users' interest is not varying much from time to time. Hence, in our application scenario (i.e., app recommendation), we re-construct graph weekly with the most up-to-date information, which takes a few hours.

\subsection{Graph Mining Algorithms}\label{sec:algo:graph:algo}

%\textcolor{red}{After the user-app graph being constructed, the second step is Graph Mining, which plays a critical role in Application Layer of PSP. In this section, we present the details of Graph Mining Algorithms, including motivation and principle. This essential part mines the potential related users to a pre-defined message which contains a set of apps.}
After the user-app graph being constructed, the second step is Graph Mining, which plays a critical role in Application Layer of PSP. In this section, we present the details of Graph Mining Algorithms, including motivation and principle. This essential part mines the potential related users to a pre-defined message which contains a set of apps.

\subsubsection{Motivation of PARW}

%At present, there are several techniques to solve the target user group discovery, which are rule-based, CF-based and graph-based. However, the rule-based method can not take advantage of collaborative information among users, because it needs a set of rules that is defined so that the accuracy rate is low and not flexible enough. The method based on collaborative filtering tends to be ineffective in real-world scenarios because it requires a great deal interactions of users and tags, which is very problematic in the real world and difficult to be fulfilled.

In this section, we present the motivation behind our graph-based algorithm. A naive way in the graph-based methods of mining user group involves two steps: firstly, label all the apps that are related to the given topic, and then select the users who have installed the apps labelled in the first step as the result. For example, suppose there is a mission to find a group of music lovers based on a graph in Figure~\ref{fig:graphexample}. Assume that, we label $A_1$ and $A_3$ as the apps that are developed for music lovers, then it is straightforward to find the users who have installed these two apps, namely, $U_1$ and $U_3$.

%In this work, we explore a approximate algorithm of \emph{PARW}, we named as \emph{A-PARW}, which is a general version of \emph{xRank} and more practical. Similar with \emph{xRank}, our advertiser just need to tell us some of the seed apps(normally 10 apps are enough) that their target users might like, then our algorithm will automatically find out the list of potential users most likely this class apps.

However, there are two serious problems with this naive graph-based approach, as stated in \cite{he2015mining}. First of all, the approach relies heavily on manual labeling of the apps, which is too expensive to manually label all the apps belonging to a specific category and limited to different business demands. Secondly the naive approach gives no ranking information which is critical for some applications. For example, for two promotion activities with different numbers but the same type of labelled apps, we cannot separate a subset from target user group with respect to the similarity with labelled apps.

In our earlier work \cite{he2015mining}, we applied PPR for the same purpose which does not rely on the labeling of all related apps and can generate the sorted user list according to their similarities with the corresponding apps in the pre-defined messages. The user-app relationship bipartite graph $\mathcal{G}$ introduced in Section~\ref{sec:algo:constr} can be viewed as representing the similarity between users by considering the overlapped apps they have installed (on the other hand, also representing the similarities between apps by considering the overlapped users who have installed these apps), therefore starting from a small set of manually-labeled \emph{seed apps}\footnote{Seed apps are labeled manually and are presented explicitly in the push message.}, the other related apps and also users who are interested in these apps can be mined from the $\mathcal{G}$ as well as their similarities to the \emph{seeds apps}.

\begin{figure}[ht]
\begin{center}
\includegraphics[width=.96\textwidth]{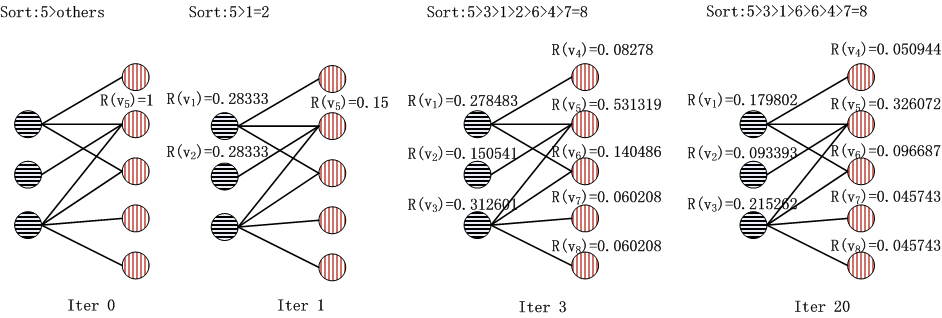}
\caption{An example of computation procedure of \emph{PPR} algorithm}
\label{fig:xrankdemo}
\end{center}
\end{figure}

However, the drawback of PPR is that it tends to neglect the relevant semi-active users even though it captures the relevant active users well. Ative users download spontaneously with high probability no matter receiving push messages or not. But there is a significant impact on semi-active users due to the fact that semi-active users do not enter Application Market very often. In Figure~\ref{fig:xrankdemo}, we present the detailed procedure of selecting target users by PPR, starting from seed app $A_2$ (namely, vertex $v_5$ ) in Figure~\ref{fig:graphexample}. Intuitively, $U_2$ (namely, vertex $v_2$) is more revelent than $U_3$ (namely, vertex $v_3$) to $A_2$, since $U_2$ only downloaded $A_2$ while $U_3$ downloaded not only $A_2$ but also $A_3, A_4, A_5$. However, we can observe from the computation procedure of PPR that the more relevant user (namely, $U_2$) is disregarded because its similarity score is taken away by the higher-degree user (namely, $U_3$).

Therefore we need an algorithm that could mine relevant semi-active users according to the given push message.  This motivates us to take a graph-based algorithm which is able to highlight community structure to generate the sorted user list according to their similarities with the corresponding push massage (namely, a few manually labelled seed apps). In the following part, we present such an algorithm, \emph{A-PARW}.

\subsubsection{A-PARW Algorithm}{\label{sec:algo:graph:algo:parw}}

As stated in the previous section, we initially label a set of apps, which are referred as \emph{seed apps} and denoted by $\mathcal{S}$. We aim to find target user group for such seed apps. A-PARW is a approximate version of Partially absorbing Random Walk Algorithm (for short, \emph{PARW} in our earlier work \cite{wu2012learning}).

We give the formulation of PARW in Equation \ref{eq:parw}.

\begin{equation}\label{eq:parw}
R^\top=I^\top \cdot (\varLambda + \mathcal{L})^{-1}\cdot \varLambda
\end{equation}

where $R$ is the rank score vector starting from $\mathcal{S}$, $\mathcal{L}$ is the Laplacian of the given graph, $\varLambda = diag(\lambda_1,\lambda_2,...,\lambda_N)$ is a diagonal matrix with $\lambda_1,\lambda_2,...,\lambda_N$ being arbitrary non-negative numbers, $I$ denotes a vector of $I(v)$ over all the vertices, $I(v)$ is the initial value of vertex $v$ initializing as:
\begin{eqnarray}\label{eq:intialvalues}
I(v)=
\begin{cases}
1/|\mathcal{S}| & \mbox{if }v\in\mathcal{S}\cr
0 & otherwise
\end{cases}
\end{eqnarray}
such that the sum of initial scores of all vertices equals to $1$.

PARW is a random walk style graph-based algorithm. In PARW, a random walk is absorbed at state $i$ with probability $p_i$, and is transferred via a random edge of state $i$ with probability $1-p_i$. We proved in~\cite{wu2012learning} that a random walk starting from a set of low conductance vertices (referred as $SP$) is most likely absorbed in $SP$ if the $\varLambda = \alpha \cdot I$ (where $I$ is an identity matrix and $\alpha$ is a small positive value). We denote the case of $\varLambda = \alpha \cdot I$ as PARW-I. One property of PARW-I is that the absorption probability varies slowly within $SP$, and drops sharply outside $SP$. This property suggests that PARW-I is able to capture the global community structure.

\begin{figure}[ht]
\begin{center}
\includegraphics[width=.96\textwidth]{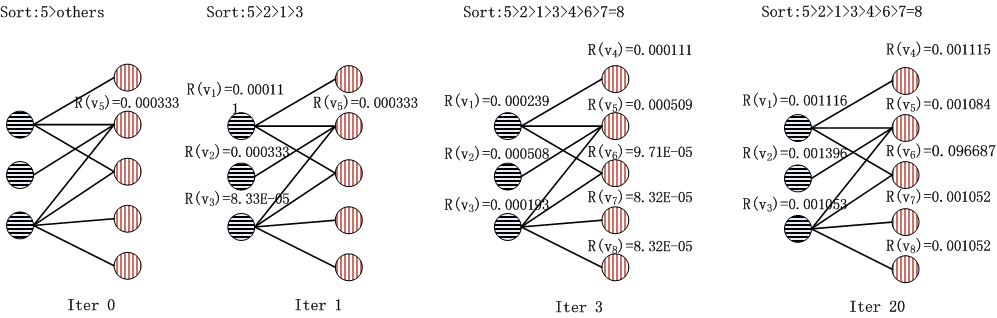}
\caption{An running example of \emph{PARW-I} algorithm}
\label{fig:scaledemo}
\end{center}
\end{figure}

In Figure~\ref{fig:scaledemo}, iteration 0 suggests that high degree vertex $v_3$ and low degree vertex $v_2$ have the same probability to get score from the seed node $v_5$. However, $v_2$ absorbs more score than $v_3$ because the user representing by $v_2$ only installs $A_2$ (namely, vertex $v_5$), which means he is more interested in $A_2$. Actually, $v_5$ transfers $\frac{1}{\lambda + d_5}=\frac{1}{\lambda +3}$ score to every neighbour, while the absorbing rate of $v_2$ is $\frac{\lambda}{\lambda+d_2}=\frac{\lambda}{\lambda+1}$, which is strictly greater than $\frac{\lambda}{\lambda+d_3}=\frac{\lambda}{\lambda+4}$ of $v_3$.

On the other hand, we showed in \cite{wu2012learning} that D mode of PARW ($\varLambda=\alpha \cdot \mathcal{D}$) is equivalent to PPR \cite{haveliwala2002topic}. Furthermore, we show that D mode of PARW is the same as PPR \cite{he2015mining}. Equation \ref{eq:xrankmatrix} is the formulation of PPR.
\begin{equation}
\label{eq:xrankmatrix}
{R^*}^\top= \alpha\cdot R^\top \cdot T + \frac{1-\alpha}{2}\cdot \frac{1}{N}\cdot {1_N}^\top + \frac{1-\alpha}{2}\cdot I^\top
\end{equation}
where $\alpha$ is a decay factor.

We show the equivalence between D mode of PARW and PPR by transforming the formulation of PPR to D mode of PARW.
\begin{align*}\label{eq:xranktransform}
{R^*}^\top &= \alpha\cdot R^\top \cdot T + (1-\alpha) \cdot (\frac{1}{2}) \cdot ({1_N}^\top/N  + I^\top)\\
& = \alpha \cdot R^\top \cdot D^{-1} \cdot W + (1-\alpha) \cdot {I^{'}}^\top\\
& =  R^\top \cdot \varLambda^{-1} \cdot W \cdot (\varLambda + D)^{-1}\cdot \varLambda  + {I^{'}}^\top \cdot (\varLambda + D)^{-1}\cdot \varLambda \\
&\because R^* \approx R\\
& \therefore  R^\top \cdot \varLambda^{-1}\cdot (\varLambda + D)= R^\top \cdot \varLambda^{-1} \cdot W + {I^{'}}^\top \\
& R^\top \cdot \varLambda^{-1}\cdot (\varLambda + \mathcal{L})  = {I^{'}}^\top \\
&R^\top  = {I^{'}}^\top \cdot (\varLambda + \mathcal{L})^{-1} \cdot \varLambda
\end{align*}
where $\varLambda = (1-\alpha)/ \alpha\cdot \mathcal{D}$, $\mathcal{L} = \mathcal{D}-\mathcal{W}$, $I^{'} = \frac{1}{2}\cdot(1_N/N + I)$, and $R^*$ equals $R$ after enough iterations.

\begin{algorithm}[htbp]
\begin{algorithmic}[1]
%\DontPrintSemicolon % Some LaTeX compilers require you to use \dontprintsemicolon instead
\Require{A start seed $s$, the regularization parameter $\varLambda=\{\lambda_1,\lambda_2,...,\lambda_n\}$, and a tolerance threshold $\gamma$ }
\Ensure{A-PARW vector $dry$}
%\Comment{update PARW value for vertex $v$}
\State Initialize $dry = 0$ and $run = \{(s, 1)\}$
\While {$run$ is \ not \ empty}
\State {pop \ a \ queue \ $run$ element \ $(i,w)$}
\State $dry_i=dry_i + \frac{\lambda_i}{\lambda_i+d_i}\cdot w$
\If {$w < \gamma\cdot d_i$}
\State {continue}
\EndIf
\For {all \ links $(i, j) \in \mathcal{\xi} $}
\If {pair $(j,s) \in run$}
\State $s=s+\frac{w}{\lambda_i+d_i}$
\Else
\State {add \ a \ new \ pair $(j, \frac{w}{\lambda_i+d_i})$ to \ $run$}
\EndIf
\EndFor
\EndWhile
\end{algorithmic}
\caption{{dry=A-PARW($s, \varLambda, \gamma$)  }\Comment \sc{approximate algorithm of parw}}
\label{algo:APARWalgo}
\end{algorithm}

However, to the best of our knowledge, there is no scalable implementation of PARW algorithm. Therefore we propose a flow-based approximation of PARW in Algorithm \ref{algo:APARWalgo} motivated by \cite{andersen2006local} and followed the iterative formula of PARW. Our algorithm maintains a pair of vectors $run$ and $dry$, starting with the trivial approximation of $dry=\vec{0}$ and $run=I$ (Line 1), and then applies a series of push operations which transfers probabilities from $run$ to $dry$ while keeps no transfer out of $dry$ . At a vertex $v_i$, a push operation transfers $\lambda_i/(\lambda_i+d_i)$ fraction of the vertex's probability from $run$ to $dry_i$ (Line 3-4), and then equally distributes the remaining $d_i/(\lambda_i+d_i)$ fraction of $run_i$ to $v_i$'s neighbours (Line 8-14). To ensure the amount of probability is moved each step and to allow us control the number of push operations, our algorithm performs push operations only on the vertices where $run_i \geq \gamma\cdot d_i$ (Line 5-7).

\subsection{Filtering Rule}\label{sec:algo:filter}

%\textcolor{red}{In real-life scenario, the target user list, which is mined through Graph Mining Algorithms, includes some users who are not suitable to sent push messages due to various reasons. Therefore we can define some practical filtering rules to filter them. For instance, we should not select users who disable the function of receiving push messages, or we may not want to send message to the users who visit Hispace Store every day, etc. This module of PSP will filter the target user list generated by the graph-based algorithm through the pre-defined rules.}
In real-life scenario, the target user list, which is mined through Graph Mining Algorithms, includes some users who are not suitable to sent push messages due to various reasons. Therefore we can define some practical filtering rules to filter them. For instance, we should not select users who disable the function of receiving push messages, or we may not want to send message to the users who visit Hispace Store every day, etc. This module of PSP will filter the target user list generated by the graph-based algorithm through the pre-defined rules.
\subsection{On-line Service}\label{sec:algo:online}

%\textcolor{red}{In the on-line part, the on-line \emph{Push Service} sends the message to the selected users. If the push service is enabled on the user's phone as well as the phone is connected to the internet, then the messages can be delivered to the user's phone and the title of the message will be shown as an alert on the notification center. Target users will receive message in their phone as shown in the left figure in Figure~\ref{fig:push}, and then users may neglect the message or click on it. After clicking the message, users will enter into the specific page of Hispace Store as shown in the middle and right figure in Figure~\ref{fig:push} or even download apps contained in the message.}

In the on-line part, the on-line \emph{Push Service} sends the message to the selected users and uses the user's feedback data to help us refine the push strategy.

Firstly, if the push service is enabled on the user's phone as well as the phone is connected to the internet, the messages can be delivered to the user's phone and the title of the message will be shown as an alert on the notification center . After that, target users will receive message in their phone as shown in Figure~\ref{fig:push-m}, and then users may neglect the message or click on it.

After clicking the message, users will enter into the specific page of Hispace Store or even download apps contained in the message. For example, when a music fans receiving an alert about music Apps on her smart phone's notification center (e.g., as the second message shown in Figure~\ref{fig:push-m}), she takes a look at music apps in the display page (e.g., as in Figure~\ref{fig:push-2}), after clicking the alert message.

In the end, we utilize users' feedback to generate market strategies and filtering rules. For example, we are more likely to send message of a music-like activity to a user who has clicked the music push activity before, and we are less likely to send a push message to a user who have never clicked any push message bofore.

\subsection{Computing Engine}\label{sec:algo:engine}

%\textcolor{red}{As presented in Section~\ref{sec:algo:constr}-Section~\ref{sec:algo:online}, there is a variety of computing tasks in Application Layer, such as raw data extraction in graph construction and target user grouping in graph mining. It is well known that data extraction is easy to parallel because computation is independent among different pieces of data. However, due to the dependencies between vertices in a graph, graph mining is very different from ordinary calculation. Therefore we choose MapReduce as general computing engine, and select more suitable graph engines for graph mining. To meet different practical needs, we select VENUS and PowerGraph as our graph engines. Specifically, VENUS is a disk-based graph engine, so it can be used when the memory is limited; on the contrary, PowerGraph is a memory-based distributed graph engine, therefore it could support more efficient calculation when memory is large enough. We implement both versions (presented in this section) in PSP for different situations.}
As presented in Section~\ref{sec:algo:constr}-Section~\ref{sec:algo:online}, there is a variety of computing tasks in Application Layer, such as raw data extraction in graph construction and target user grouping in graph mining. It is well known that data extraction is easy to parallel because computation is independent among different pieces of data. However, due to the dependencies between vertices in a graph, graph mining is very different from ordinary calculation. Therefore we choose MapReduce as general computing engine, and select more suitable graph engines for graph mining. To meet different practical needs, we select VENUS and PowerGraph as our graph engines. Specifically, VENUS is a disk-based graph engine, so it can be used when the memory is limited; on the contrary, PowerGraph is a memory-based distributed graph engine, therefore it could support more efficient calculation when memory is large enough. We implement both versions (presented in this section) in PSP for different situations.

\subsubsection{Implementation of PARW}\label{sec:algo:engine:imple}

We designed A-PARW based on VENUS for single machine and PowerGraph for a distributed cluster. PowerGraph is a memory-based graph computing engine which is very efficient with sufficient memory, while VENUS is a disk-based engine which needs only limited memory but not very efficient compared to PowerGraph. In this section, we give the implementation of the $update function$ on VENUS and the $apply function$ on PowerGraph for each vertex in Algorithm \ref{algo:APARWupdate-V} and Algorithm \ref{algo:APARWupdate-PG}, respectively.

\begin{algorithm}
\begin{algorithmic}[1]
%\DontPrintSemicolon % Some LaTeX compilers require you to use \dontprintsemicolon instead
\Require{vertex $v$, initial score [$v.run$,$v.dry$], stop threshold \emph{$\gamma$}, the regularization parameter $\varLambda=\{\lambda_1,\lambda_2,...,\lambda_n\}$ }
\Ensure{updated value of $v$}
%\Comment{update PARW value for vertex $v$}

 \State  $v.dry += \frac{\lambda_v}{\lambda_v + d_v} \times v.run\;$

\For{$u \in v.neighbour$}
%{
   \If{$u.run > \gamma\cdot d_u$ }
%   {
 \State  $v.run += \frac{1}{\lambda_u + d_u} \times u.run\;$
%   }
   \EndIf
%}
\EndFor
\State \Return {v.runflow,v.dryflow.}
\end{algorithmic}
\caption{{\sc{function} \emph{update} on \sc{venus}}}
\label{algo:APARWupdate-V}
\end{algorithm}

As shown in Algorithm \ref{algo:APARWupdate-V}, the A-PARW algorithm is implemented in the $update$ function of VENUS. Firstly, we initialize the $run=I_i$ before the first iteration. Then we transfer $\frac{\lambda_v}{\lambda_v + d_v}$ from $run$ to $dry$ (Line 1). For each iteration, if the $run$ of any neighbour $u$ of $v$ is larger than $\gamma\cdot d_u$, we transfer $\frac{1}{\lambda_u + d_u}$ from $run_u$ to $run_v$ (Line 2-5).

Different from $update$ function on VENUS, A-PARW on PowerGraph (shown in Algorithm~\ref{algo:APARWupdate-PG}) executes procedure $init$, $apply$ and $scatter$ sequentially. $init$ initializes $run$ at first iteration and updates it by the summed up the map value from last remaining scatter (Line 3-11). After that, it updates $run$ and $dry$ of vertex $v$ in procedure $apply$ (Line 12-17) and procedure $scatter$ pushes remaining $run$ value to the neighbours if the remnant is larger than $\gamma$ (Line 18-24).

To end this section, we discuss how we choose the parameters of the algorithms. As discussed in~\cite{wu2012learning}, the smaller regularization parameter $\alpha$ we selected, the better of performance PARW-I achieved. However, due to the large scale of our User-APP graph, extremely small $\alpha$ would bring computing error, resulted by overflow. Therefore, we choose a relative small $\alpha=0.01$ to obtain both better performance and computation reliability.

%\textcolor{red}{In A-PARW, $dry$ is almost equivalent to PARW results when the $run$ is less than $\gamma$ after $n$ iterations when ${\frac{\bar{d}}{\alpha + \bar{d}}}^n\approx \gamma$ , where $\bar{d}$ is the average degree. It can be calculated that $n$ is a particularly large number when $\alpha$ and $\gamma$ are both small. Luckily, we found that in the actual push service, $run$ does not need to flow everything to $dry$, only one thousandth or even less would be enough to get good approximate result. As a result, we set $\gamma$ to be $10^{-12}$ and select a limited number of iterations (i.e., 20) as A-PARW-I's stop condition. It is demonstrated to be good enough in our scenario.}
In A-PARW, $dry$ is almost equivalent to PARW results when the $run$ is less than $\gamma$ after $n$ iterations when ${\frac{\bar{d}}{\alpha + \bar{d}}}^n\approx \gamma$ , where $\bar{d}$ is the average degree. It can be calculated that $n$ is a particularly large number when $\alpha$ and $\gamma$ are both small. Luckily, we found that in the actual push service, $run$ doesn't have to flow everything to $dry$, only one thousandth or even less would be enough to get good approximate result. As a result, we set $\gamma$ to be $10^{-8}$ and select a limited number of iterations (i.e., 20) as A-PARW-I's stop condition. It is demonstrated to be good enough in our scenario.
\begin{algorithm}
\begin{algorithmic}[1]
\State \textbf{Data}: $run\_map \ is \ a \ map \ and \ run\_map[u] \ represents \ runflow_u(v)$
\State \textbf{Data}: $dry\_map \ is\ a\ map\ and\ dry\_map[u]\ represents\ dryflow_u(v)$
%\DontPrintSemicolon % Some LaTeX compilers require you to use \dontprintsemicolon instead
\Procedure{init}{$v$}
\If {\emph{first iteration}}
\If {$v \in S$}
\State $run\_map[v] \gets I(v)$
\EndIf
\Else
\State $run\_map \gets prev\_run$
\EndIf
\EndProcedure
\Procedure{apply}{$v$}
\ForAll{$\emph{key-value \ pair} (w,t)> \in run\_map$}
\State $dry\_map[w] \gets dry\_map + \frac{\lambda_v}{\lambda_v + d_v}\times t$
\State $run\_map[w] \gets \frac{1}{\lambda_v + d_v}\times t$
\EndFor
\EndProcedure
\Procedure{scatter}{$v$}
\ForAll{$\emph{key-value \ pair} (w,t)> \in run\_map$}
\If {$w \geq \gamma$}
\State $send \ run\_map \ to \ vertex \ u\ where (v,u)\in \mathcal{\xi} $
\EndIf
\EndFor
\EndProcedure
\end{algorithmic}
\caption{{\sc a-parw on powergraph}}
\label{algo:APARWupdate-PG}
\end{algorithm}

\subsubsection{Comparison of PowerGraph and VENUS}\label{sec:algo:engine:compare}

In this section, we prepare the efficiency performance of A-PARW-I, implementing on PowerGraph and VENUS respectively. When the push activity is not very urgent and the memory resource is limited, we can choose the VENUS implementation. However, if the push activity is urgent, and even worse we have to handle tens or hundreds of push activities in a short period, it is better to choose the PowerGraph implementation.

In order to compare the efficiency of the two versions of implementation, we run experiments on twitter-graph \cite{kwak2010twitter}, which contains 41,652,230 vertices and 1,468,364,884 edges. The parameters are set to be the same as in the real push activities (as stated in the previous section). We run the experiments on a server with two eight-core Intel Xeon E5-2650 2.60 GHz processors, 377 GB RAM, and 20 TB hard disk, with Ubuntu 14.04. To compare fairly, we disable the distribute-mode of PowerGraph. The evaluation results are presented in Table~\ref{tb:runtime}. To process one push activity, PowerGraph needs only 6.79 seconds, while VENUS needs around 40 minutes. For 10 push activities, PowerGraph uses around 13 seconds, while VENUS have to run more than 6 hours. If we need 100 push activities, PowerGraph is able to return the results in less then 50 seconds, and we do not test it on VENUS as it needs several days to get the precise timing. Since VENUS runs the push activities one by one, therefore we can expect VENUS to finish 100 push activities in two or three days.

\begin{table}[h]
\centering
\caption{Running Time (in seconds) of A-PARW-I on PowerGraph and VENUS}
\begin{tabular}{|c|c|c|c|} \hline
No. of push activities        & 1 & 10 & 100 \\ \hline
PowerGraph   & 6.79 & 13.19 & 46.69\\
VENUS    & 2307.00 & 22588.90 & N.A.\\
\hline\end{tabular}
\label{tb:runtime}
\end{table}

\section{Experiment}\label{sec:expe}

In this section, we firstly describe two datasets used in our experiments as well as evaluation metrics in Section~\ref{sec:expe:data-set}. Next, we compare the experimental results of A-PARW-I and PPR on the two datasets, in Section~\ref{sec:expe:simul} and Section~\ref{sec:expe:real} respectively.
%Lastly, we discuss the affect of changing the number of iterations of \emph{PARW} algorithm in Section~\ref{sec:exp:iteration}.

\subsection{Data Set Description and Experiment Setting}\label{sec:expe:data-set}

We evaluated A-PARW algorithms on two datasets. One dataset is MovieLens \footnote{\url{http://grouplens.org/datasets/movielens/}}, collected by GroupLens Research. While the other dataset is collected from Hispace Store in mainland China  \footnote{\url{http://app.hicloud.com/}}. The difference between A-PARW-I and PPR is verified on both datasets, while evaluation on the real-life datasets furthermore demonstrated the effectiveness of A-PARW-I. To start with, in this subsection, we will provide a concrete description of these two datasets, as well as that of the experiment setting.

\subsubsection{Public Dataset and Experiment Setting}

As we discussed in Section~\ref{sec:algo}, PARW-I is able to identify more semi-active users than
PPR because PARW-I considers global community structure, rather than inclining high-degree nodes as PPR does. In other words, PPR prefers to rank high-degree nodes on the top, compared to PARW-I.

One way to verify this property is comparing degree distributions of user-nodes, sorting by the two algorithms. In order to illustrate this property, we ran PPR and A-PARW-I on MovieLens dataset and compared degree trends of their ranking lists.

\begin{table}[h]
\centering
\caption{Statistics Description of MovieLens Dataset}
\begin{tabular}{|c|c|c|c|c|} \hline
Item        & MovieLens 100k & MovieLens 1m & MovieLens 10m & MovieLens 20m\\ \hline
No. Users   & 943 & 6,040& 71,567 & 138,493\\
No. Movies    & 1,682& 3,883& 10,681 & 27,279\\
No. ratings   & 100,000& 1,000,209& 10,000,054 & 20,000,263\\
\hline\end{tabular}
\label{tb:dataset-movie}
\end{table}

Different copies of the MovieLens dataset were available online. They are collected over different periods, resulting in different sizes. MovieLens dataset includes seven such copies and we used four of them, namely MovieLens 100K, MovieLens 1M, MovieLens 10M and MovieLens 20M. In Table~\ref{tb:dataset-movie}, the statistics description of the copies that we used is presented.

To construct a user-movie bi-partite graph from MovieLens dataset following the method in Section~\ref{sec:algo}, we separated users and movies to two kinds of nodes and there exists an edge $(i,j)$ if user $i$ rates movie $j$. Next we randomly selected 1 (or 10 or 100) movie-nodes as seed nodes. Starting with the selected seed nodes, we ran PPR and A-PARW-I to get the degree distribution of the user nodes ranking by the two algorithms, respectively. Note that each experiment we repeated 50 times to eliminate the affect of random choices of seed nodes. The result will be presented in Section~\ref{sec:expe:simul}. Also note that for experiments on both public dataset and real-life dataset, we set the number of iterations of A-PARW algorithms to be 20.

\subsubsection{Real-Life Dataset and Experiment Setting}

Besides running the algorithms on a public dataset (i.e., MovieLens), we also performed experiments on a commercial dataset from Hispace Store in mainland China (we refer to this dataset as \emph{APPData}). According to IDC \footnote{ \url{http://www.idc.com/getdoc.jsp?containerId=prCHE41676816}}, Huawei was the third largest global smartphone manufacturer in Q3 2016. The log servers of Hispace Store record the information about users downloading apps at every and each moment. We select the complete downloading log from $2015/03/01$ to $2015/08/31$. Table \ref{tb:dataset} describes statistics of the dataset. We constructed a bi-partite graph on the basis of \emph{APPData}, in a similar way as we did on the MovieLens dataset: there exits an edge between user $i$ and app $j$ if user $i$ downloaded app $j$ during the period that we considered.

\begin{table}[h]
\centering
\caption{Dataset from a commercial App Market}
\begin{tabular}{|l|r|} \hline
Item        & Value \\ \hline
No. Users   & 96,324,654\\
No. Apps    & 487,649\\
No. Edges   & 1,778,160,959\\
Duration    & March.1 - August.31 2015 \\
\hline\end{tabular}
\label{tb:dataset}
\end{table}

We implemented the two A-PARW algorithms (namely A-PARW-I and PPR) on the VENUS \cite{cheng2015venus}. We firstly conducted experiments to evaluate the effectiveness of PARW-I, then we also perform experiments to verify the property of A-PARW-I and PPR (as discussed in Section~\ref{sec:algo}).

Our experiments on \emph{APPData} include both off-line and on-line evaluation. Firstly, we describe the off-line evaluation method. For off-line evaluation purpose, we collected users' feedback of nine push activities from Hispace Store, for which the selected user lists were generated by PPR. The topics of these push activities are present in Table \ref{tb:off-line topic}. We referred the users' feedback as the ground truth to compare the effectiveness of PPR and A-PARW-I. After receiving a push message of apps, an interested user may click on it, or even download the recommended apps. Therefore, we can distinguish the cases of clicking and downloading, as follows: we refer a sample that the user clicked (downloaded, respectively) the recommended advertisement as positive, and a sample of the opposite case as negative. We ran PPR and A-PARW-I on \emph{APPData}, and got top 1 million users as well as their ranking scores, respectively. In the off-line experiment, we adopt AUC (Area under ROC Curve) as the evaluation metric.

\begin{table}[h]
\centering
\caption{The description of off-line topics}
\begin{tabular}{|c|c|c|c|c|c|c|c|c|c|} \hline
Item & 1 & 2 & 3 & 4 & 5 & 6 & 7 & 8 & 9\\ \hline
Name  & music & camera & instrument & ticket & listen book & travel & goodnight & read & internet\\
\hline\end{tabular}
\label{tb:off-line topic}
\end{table}

In the on-line evaluation, we sent push messages to the same number of top users ranked by A-PARW-I and PPR, respectively. Making sure to receive the feedback from the majority of the users after two days, we compared the number of users who clicked (or downloaded) the recommended apps, across the two sets of users picked by the two algorithms.  We use CTR (click-through ratio)\cite{he2015mining} and also DTR (download-through ratio) as the online evaluation metrics. In our scenario, CTR and DTR are defined as follows:
\begin{align}
CTR &= \frac{\mbox{No. of users who ever clicked ads}}{\mbox{No. of users who ever received ads}}
\end{align}
\begin{align}
DTR &= \frac{\mbox{No. of users who ever downloaded ads}}{\mbox{No. of users who ever received ads}}
\end{align}

\subsection{Evaluation on Public Dataset}\label{sec:expe:simul}

Recall that in Section~\ref{sec:algo}, we stated that compared to A-PARW-I, PPR prefers to rank nodes with high degree on the top. In this section, we verify this statement empirically, i.e., we ran PPR and A-PARW-I on MovieLens dataset and compared degree distribution of their ranking list. More specifically, we random selected $x$ movie-nodes as seed nodes (we have different cases of $x=1, x=10, x=100$), then we ran PPR and A-PARW-I to generate user lists by ranking the scores, and lastly we are able to get the degree distribution of the user lists. For the ease of presentation, we integrate the user-nodes in the ordered user lists into 2,000 equal-size buckets (the size of each bucket is $\lfloor{\frac{Number\_of\_Nodes}{2000}}\rfloor$). We define the degree of each bucket as the average degree of the users included in this bucket.

\begin{figure}
\subfloat[MovieLens 100k with 1 seed]{
\label{100k-1}
\includegraphics[width=.32\textwidth]{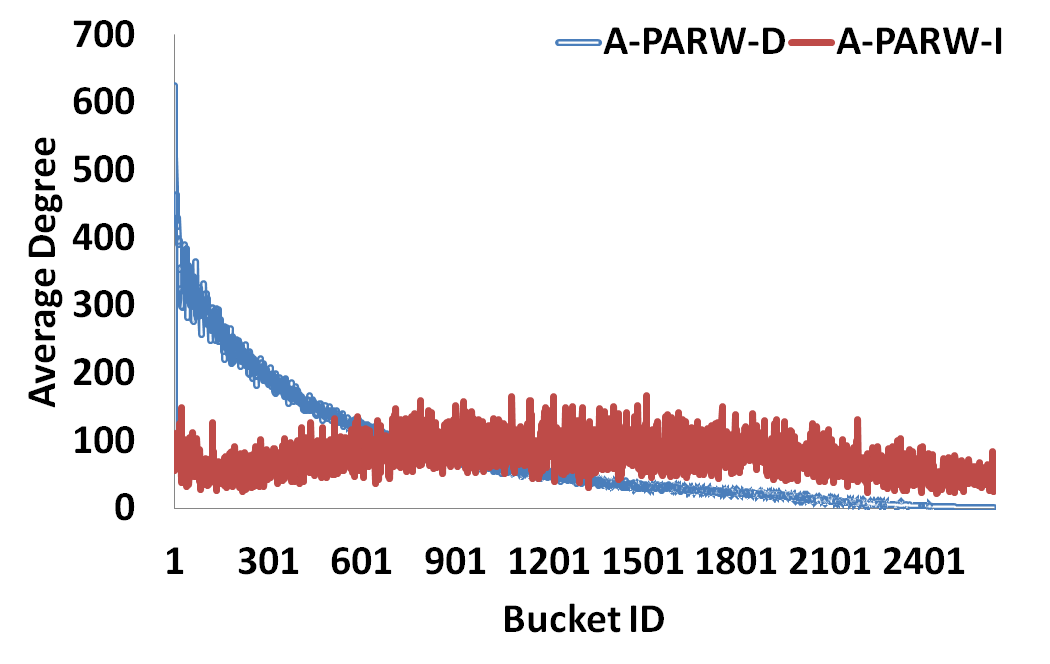}}
\subfloat[MovieLens 100k with 10 seeds]{
\label{100k-10}
\includegraphics[width=.32\textwidth]{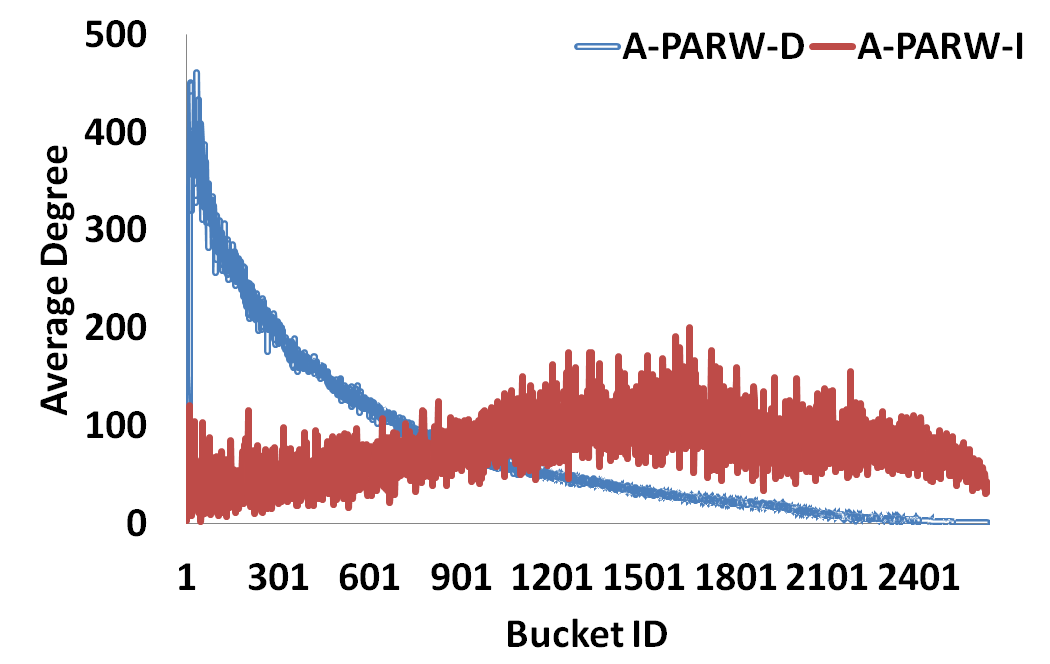}}
\subfloat[MovieLens 100k with 100 seeds]{
\label{100k-100}
\includegraphics[width=.32\textwidth]{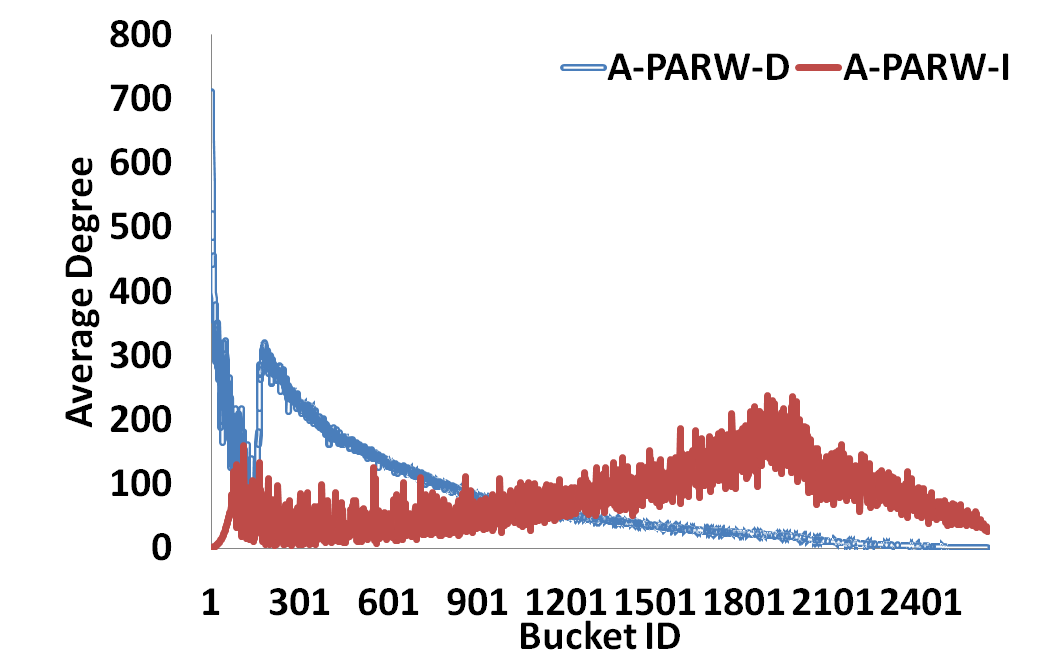}}

\subfloat[MovieLens 1m with 1 seed]{
\label{1m-1}
\includegraphics[width=.32\textwidth]{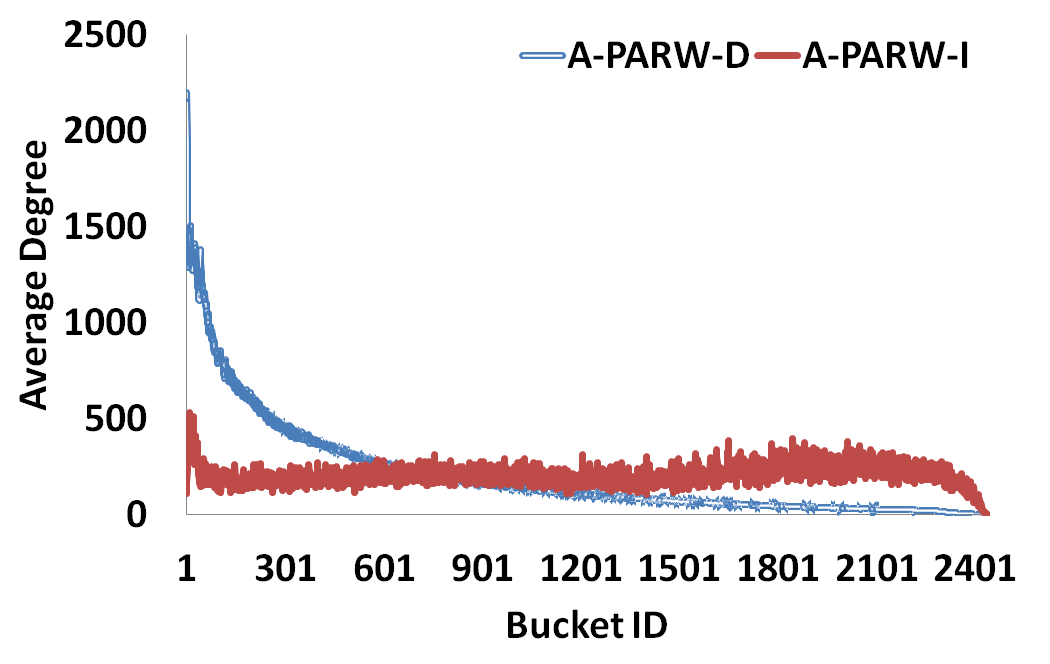}}
\subfloat[MovieLens 1m with 10 seeds]{
\label{1m-10}
\includegraphics[width=.32\textwidth]{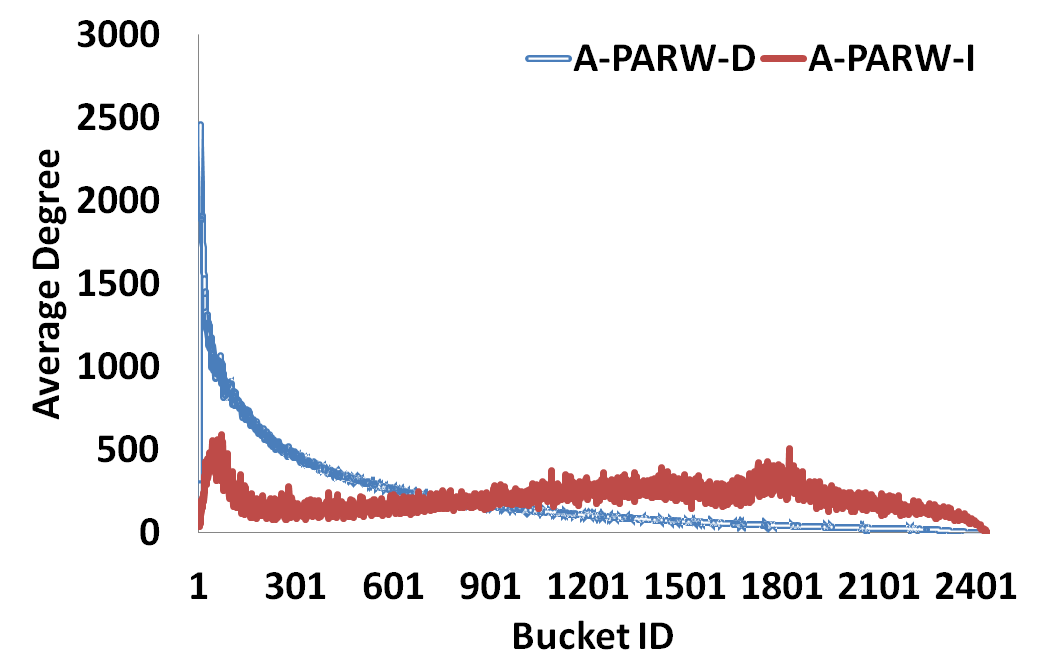}}
\subfloat[MovieLens 1m with 100 seeds]{
\label{1m-100}
\includegraphics[width=.32\textwidth]{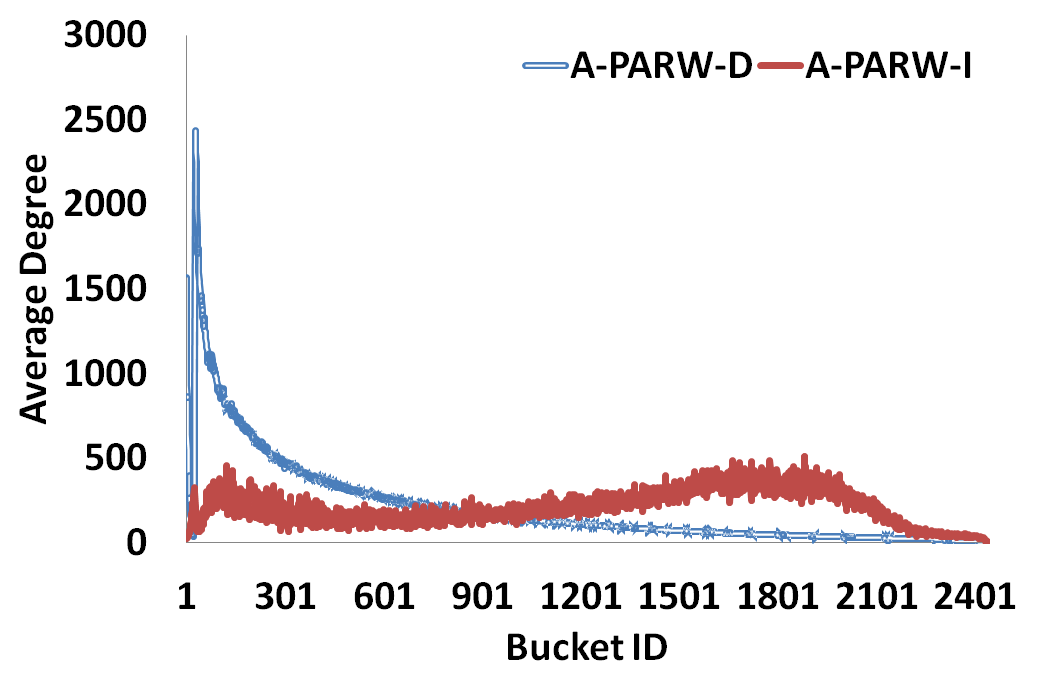}}

\subfloat[MovieLens 10m with 1 seed]{
\label{10m-1}
\includegraphics[width=.33\textwidth]{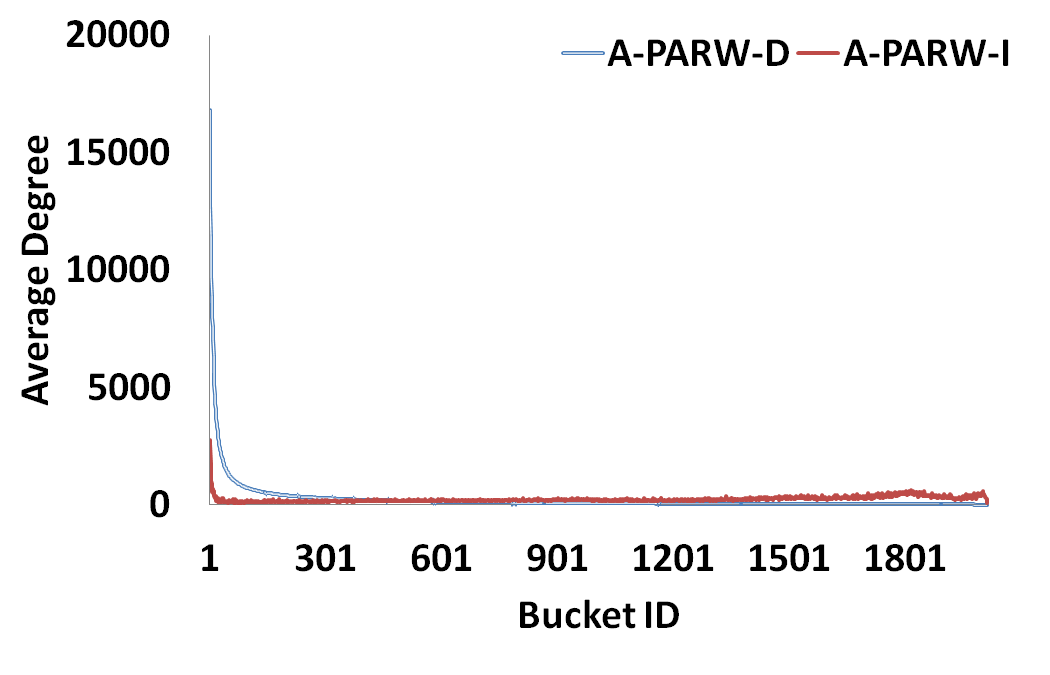}}
\subfloat[MovieLens 10m with 10 seeds]{
\label{10m-10}
\includegraphics[width=.33\textwidth]{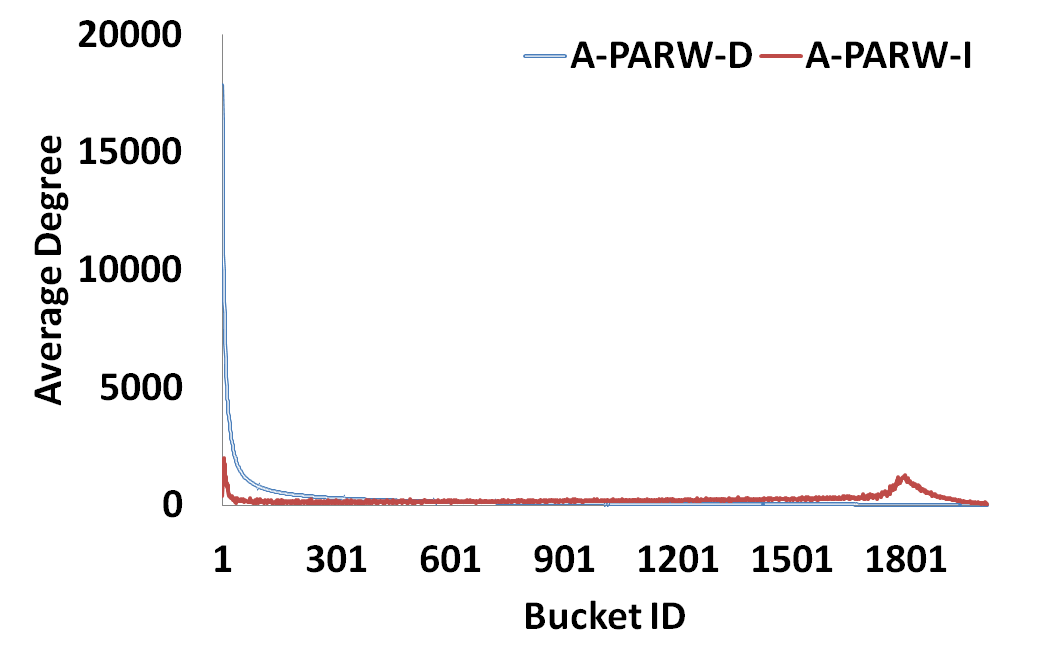}}
\subfloat[MovieLens 10m with 100 seeds]{
\label{10m-100}
\includegraphics[width=.33\textwidth]{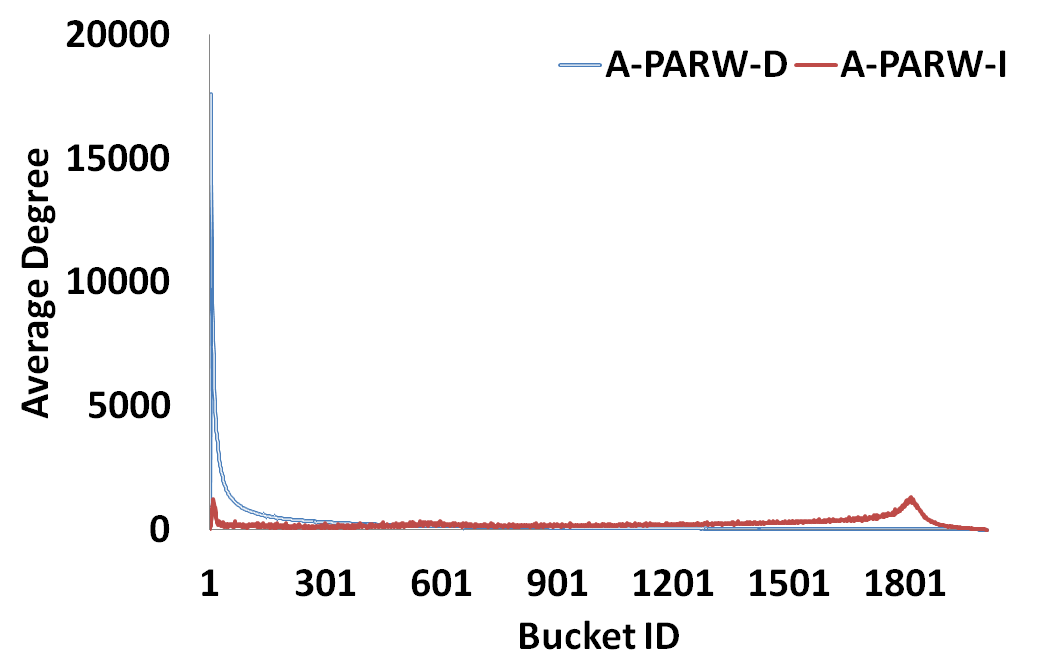}}\\

\subfloat[Zoom in Figure~\ref{10m-1}]{
\label{10m-1-p}
\includegraphics[width=.33\textwidth]{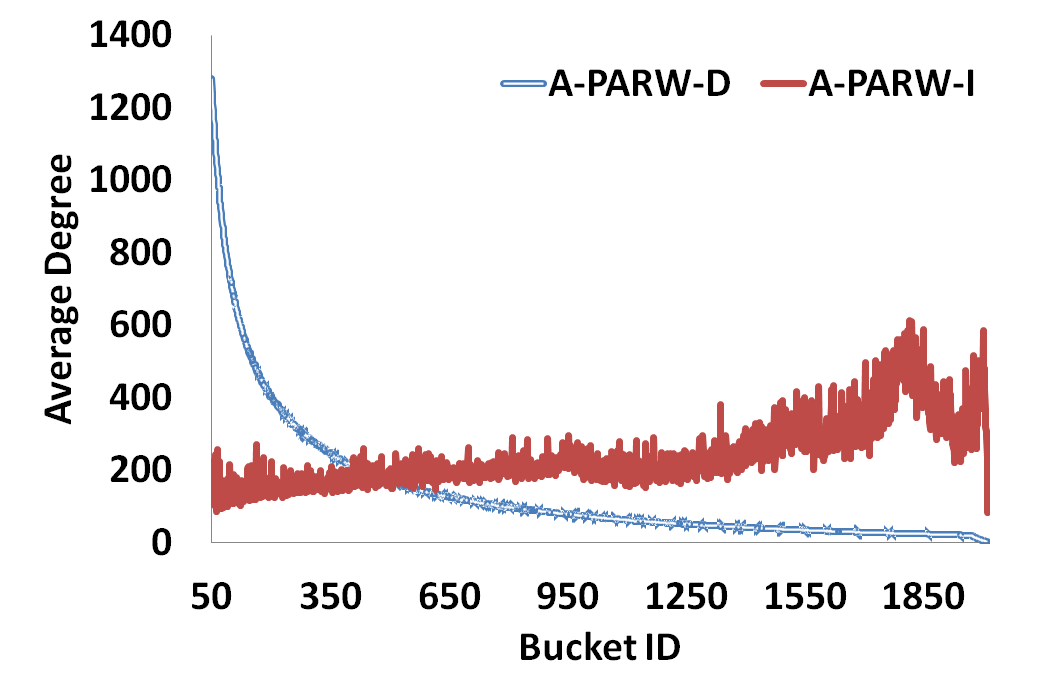}}
\subfloat[Zoom in Figure~\ref{10m-10}]{
\label{10m-10-p}
\includegraphics[width=.33\textwidth]{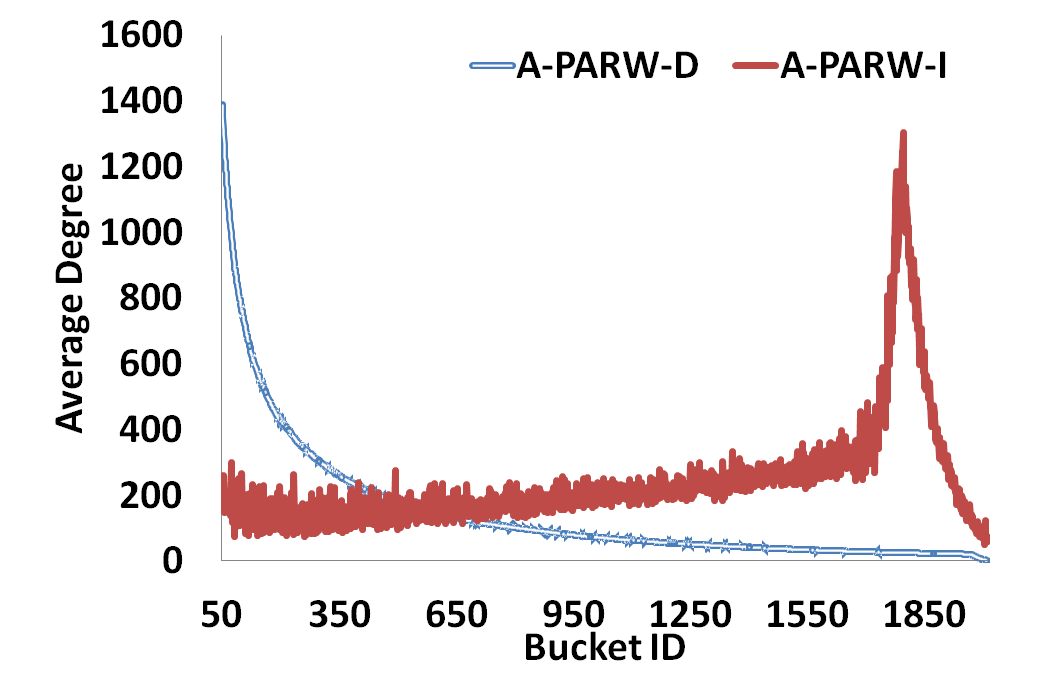}}
\subfloat[Zoom in Figure~\ref{10m-100}]{
\label{10m-100-p}
\includegraphics[width=.33\textwidth]{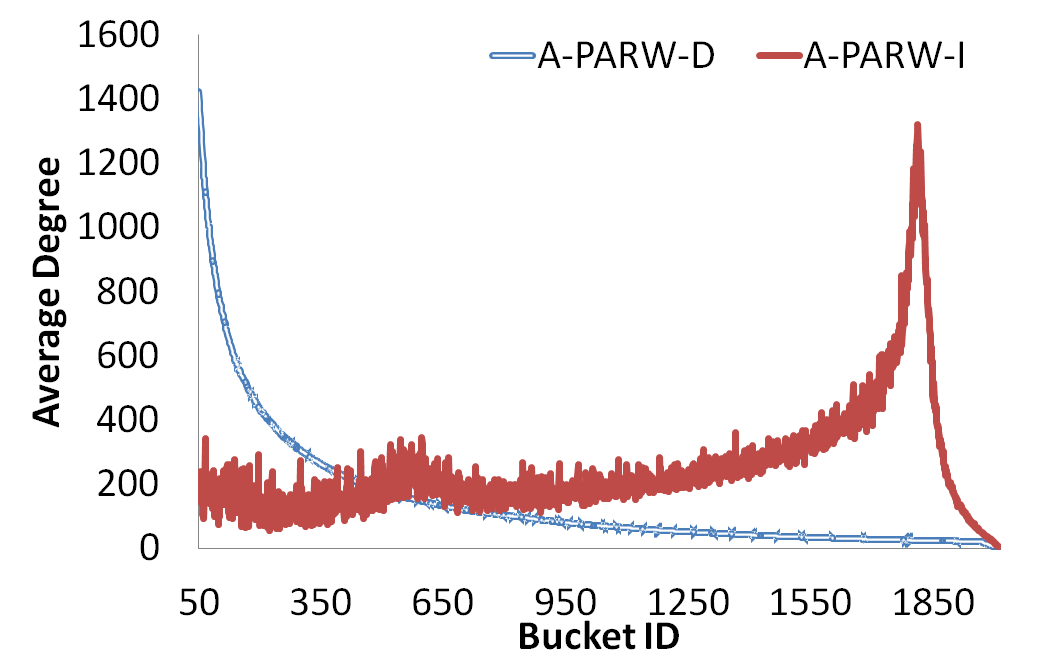}}

\subfloat[MovieLens 20m with 1 seed]{
\label{20m-1}
\includegraphics[width=.33\textwidth]{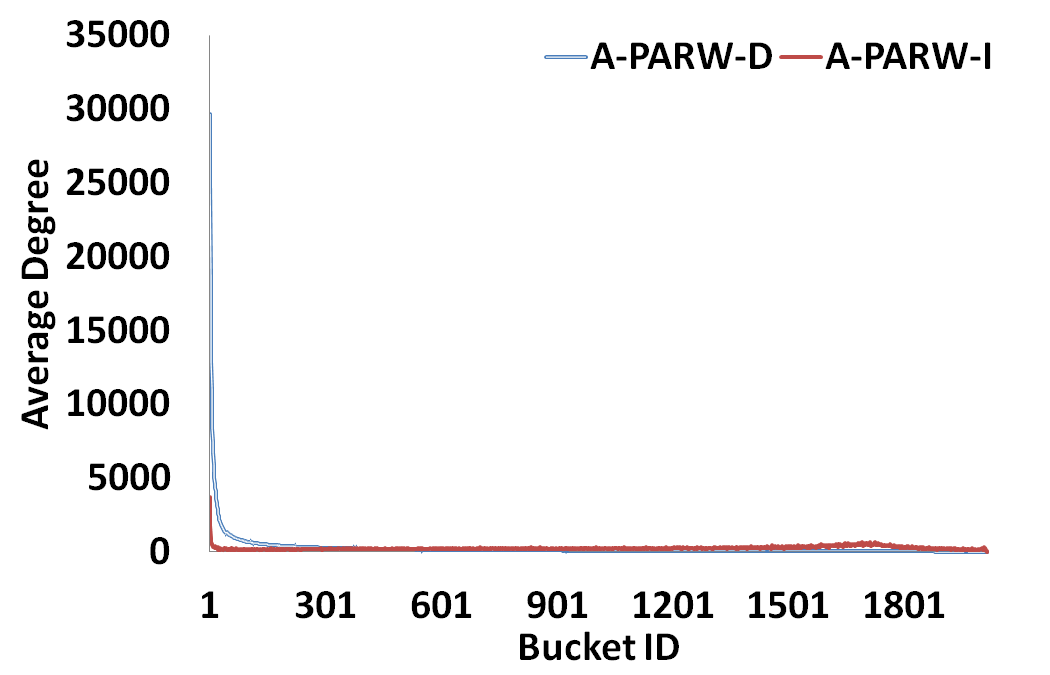}}
\subfloat[MovieLens 20m with 10 seeds]{
\label{20m-10}
\includegraphics[width=.33\textwidth]{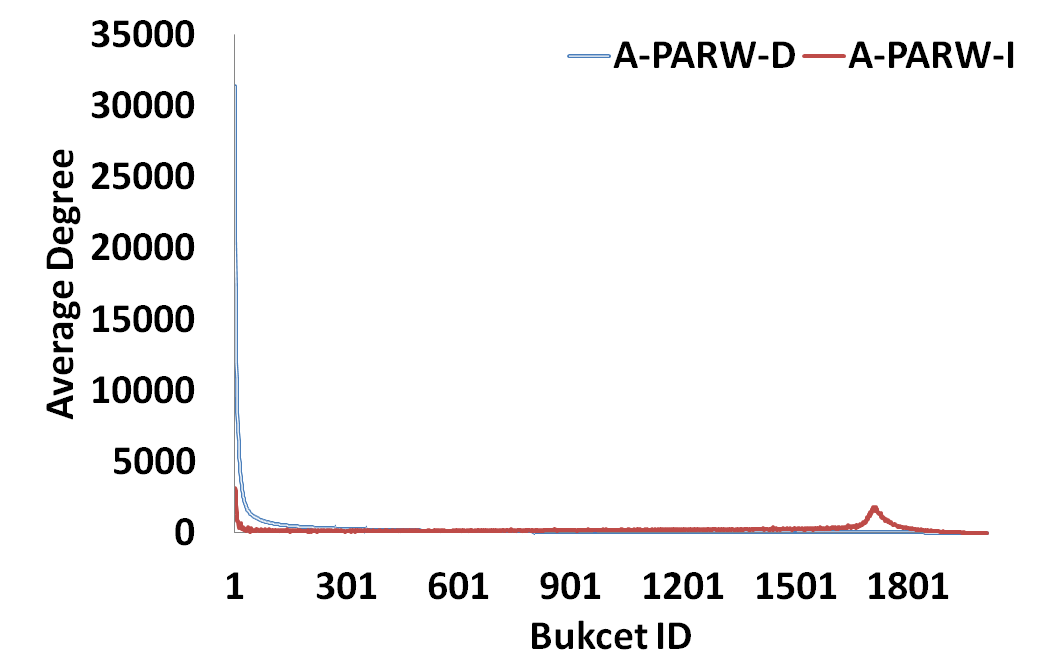}}
\subfloat[MovieLens 20m with 100 seeds]{
\label{20m-100}
\includegraphics[width=.33\textwidth]{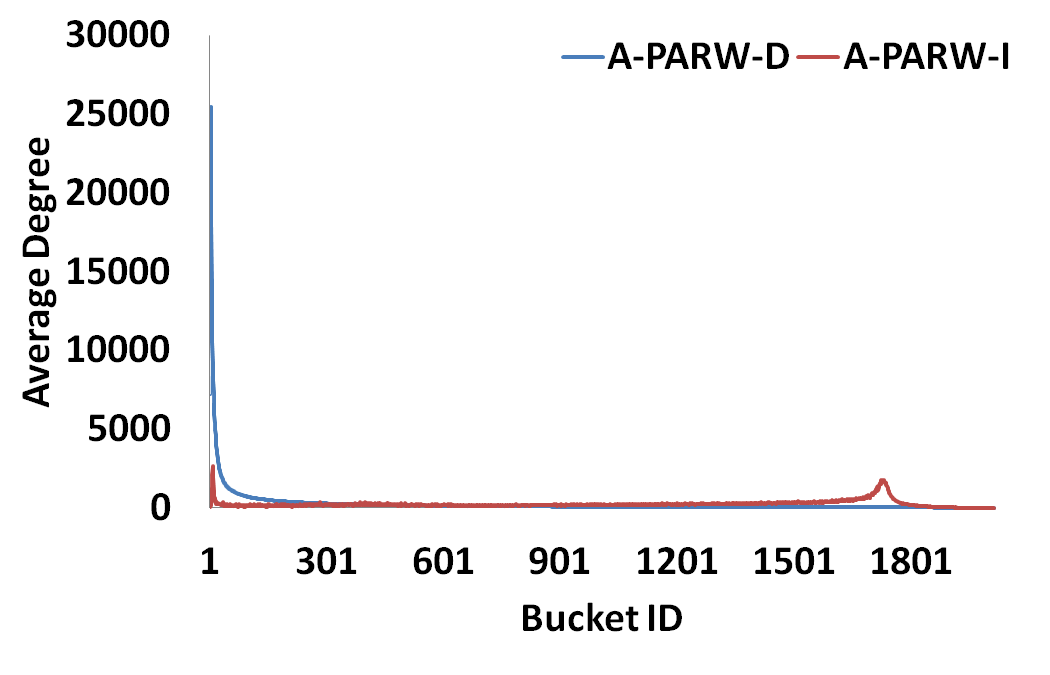}}\\

\subfloat[Zoom in Figure~\ref{20m-1}]{
\label{20m-1-p}
\includegraphics[width=.33\textwidth]{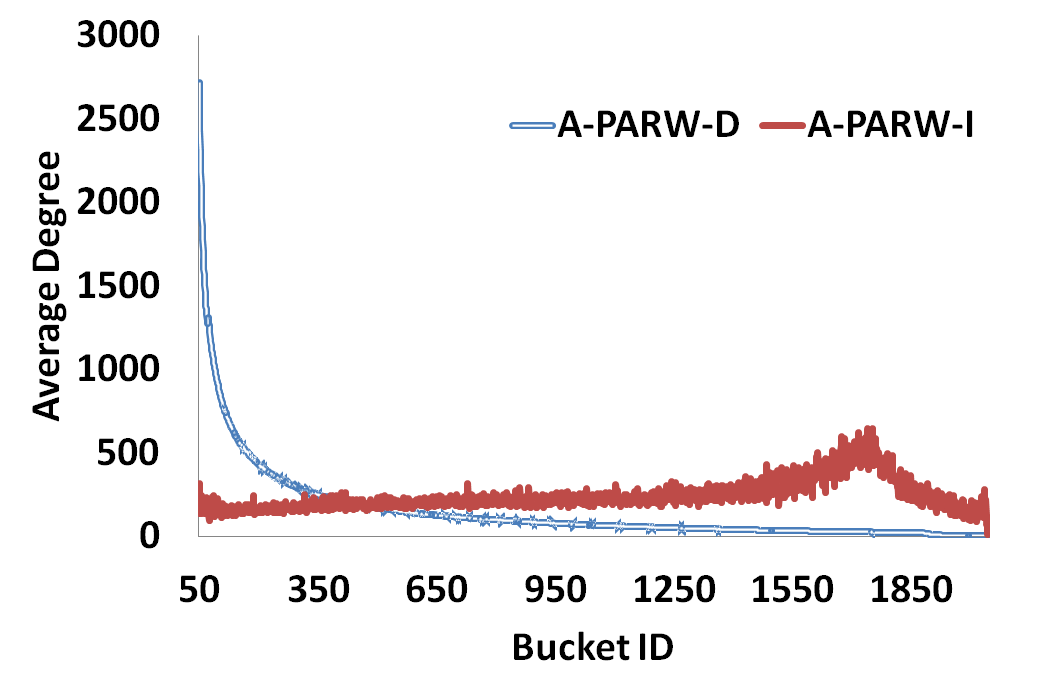}}
\subfloat[Zoom in Figure~\ref{20m-10}]{
\label{20m-10-p}
\includegraphics[width=.33\textwidth]{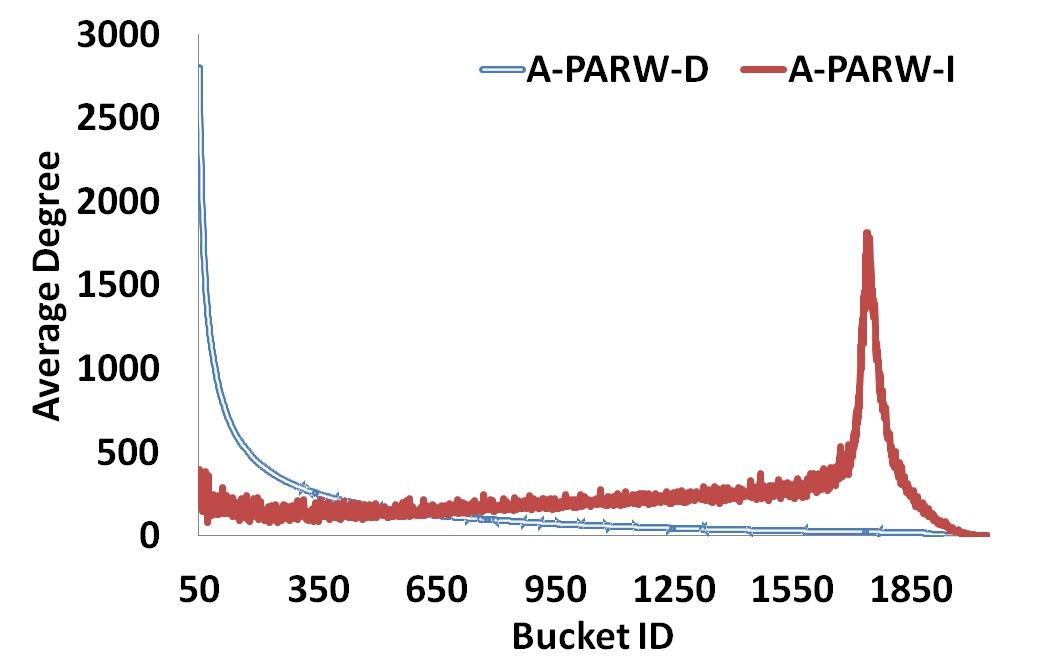}}
\subfloat[Zoom in Figure~\ref{20m-100}]{
\label{20m-100-p}
\includegraphics[width=.33\textwidth]{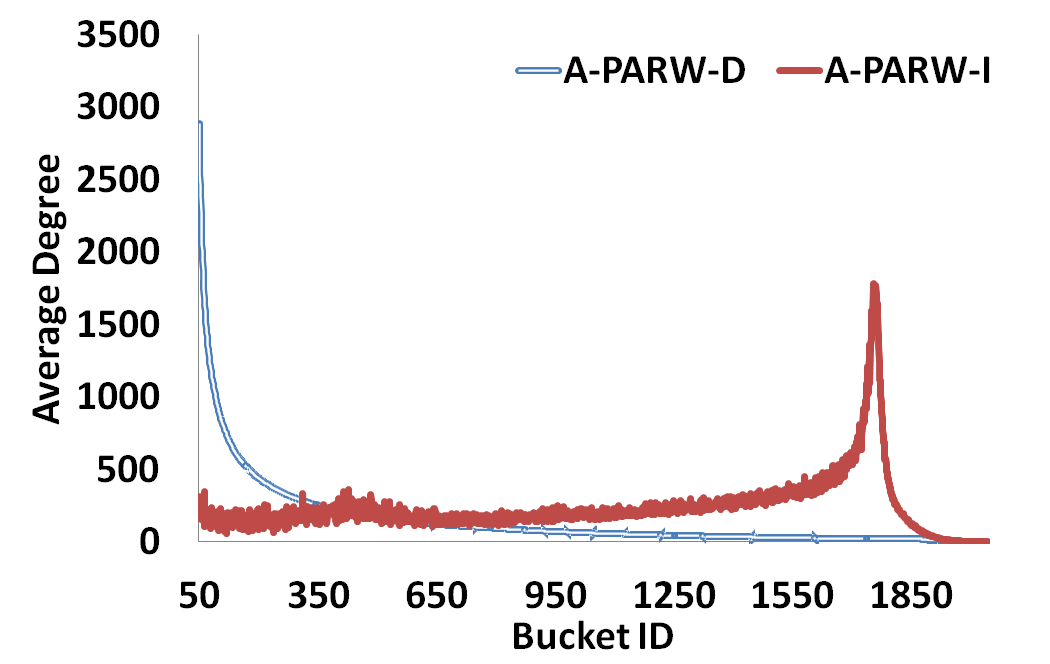}}
\caption{Evaluation Results on MovieLens}
\label{fig:MovieLens}
\end{figure}

In Figure~\ref{fig:MovieLens}, we present the degree distribution of the results of PPR and A-PARW-I on different copies of MovieLens dataset. In Figure~\ref{100k-1}, Figure~\ref{100k-10} and Figure~\ref{100k-100}, we present the evaluation result on MovieLens 100k (refer to Table~\ref{tb:dataset-movie}), and the number of seed nodes is 1, 10 and 100, respectively. From Figure~\ref{1m-1} to Figure~\ref{1m-100}, from Figure~\ref{10m-1} to Figure~\ref{10m-100} and from Figure~\ref{20m-1} to Figure~\ref{20m-100}, we show the evaluation results on MovieLens 1m, MovieLens 10m and MovieLens 20m, respectively. The overall distribution on MovieLens 10m (i.e., from Figure~\ref{10m-1} to Figure~\ref{10m-100}) and MovieLens 20m (i.e., from Figure~\ref{20m-1} to Figure~\ref{20m-100}) is not clear for us to observe interesting findings. Therefore, we zoom into the tails of curves, as shown from Figure~\ref{10m-1-p} to Figure~\ref{10m-100-p} and from Figure~\ref{20m-1-p} to Figure~\ref{20m-100-p}. Every figure reports the corresponding result by repeating the experiment 50 times, in order to eliminate the affect of random choices of seed nodes. In each figure, x-axis represents the bucket identifier and y-axis the degree of the corresponding bucket. The blue curve is the degree distribution of the users list generated by PPR, while the red curve represents that of A-PARW-I.

It can be observed from Figure \ref{fig:MovieLens} that the degree distribution of users list by PPR (the blue curve) is higher than A-PARW-I (the red curve) at beginning, and then blue curve decreases rapidly and at last the blue curve lays below the red one. In addition, the degree distribution of A-PARW-I (the red curve) is more steady and stable except for a few of buckets near the tail. The reason for this exception is that the scores of high degree irrelevant nodes could be larger than the scores of low degree irrelevant nodes.

This observation is consistent with the discussion in Section~\ref{sec:algo}. It can be further implied that PPR prefers to higher degree user-nodes which results in losing semi-active users (whose degree is not very high). In contrast, A-PARW-I is suitable to capture semi-active users because it prefers community structure, rather than high degree nodes.

\subsection{Evaluation on Real-Life Dataset}\label{sec:expe:real}

To evaluate the effectiveness of A-PARW-I in a real-world recommender system, we compare the performance between PPR and A-PARW-I, by running off-line experiment on 9 push activities and on-line experiment on 2 push activities through Hispace Store's push service. Note that the effectiveness of PPR is demonstrated by our previous work \cite{he2015mining}.

\subsubsection{Off-line Evaluation}

In off-line experiment, we compare the performance of A-PARW-I with PPR by comparing the ranking accuracy of their result list (we get top 1 million users from their result list). We use AUC as the evaluation metrics. The ground truth is determined by user feedback (click, download, or nothing), collected from log server of Hispace Store.

\begin{figure}
\begin{center}
\includegraphics[width=.55\textwidth]{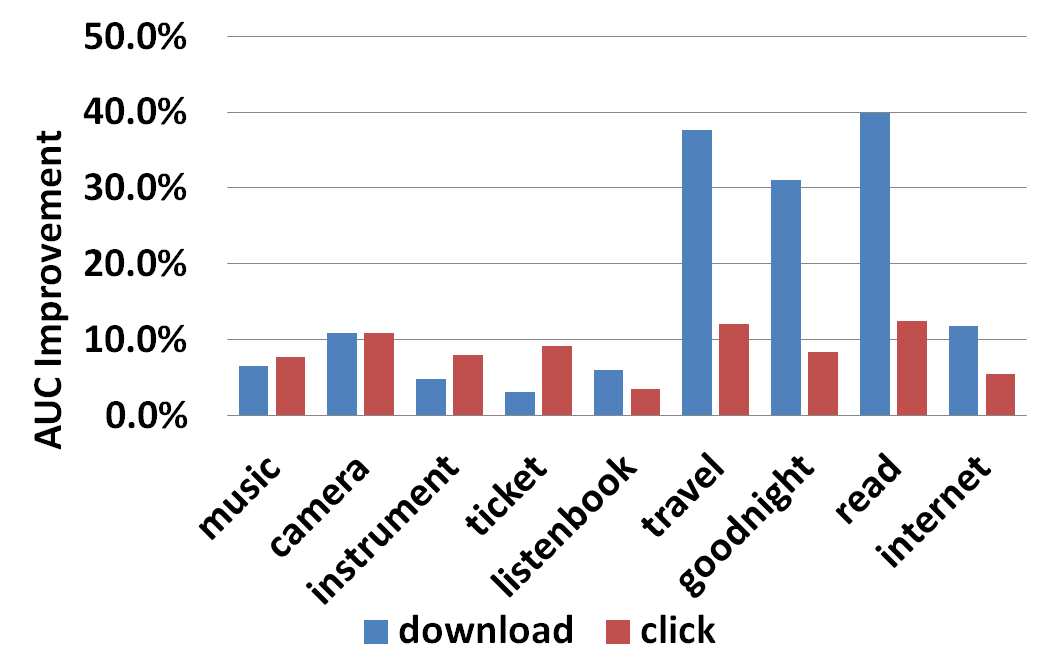}
\caption{The AUC improvement of A-PARW-I over PPR}
\label{fig:offline}
\end{center}
\end{figure}

In Figure~\ref{fig:offline}, we present click AUC and download AUC improvement of A-PARW-I over PPR on the 9 push activities. In each figure, x-axis represents topic name of the push activity while y-axis is AUC improvement.

As can be observed, the performance of A-PARW-I is better than PPR at all the 9 push activities, on both click and download cases. Moreover, the improvement of A-PARW-I over PPR is more significant on download case, compared to the click case. The improvement of A-PARW-I comes from the fact that A-PARW-I pays more attention to graph community information while PPR over considers degree of nodes in the graph; in other words A-PARW-I could find more semi-active user-nodes, which are of low degree but highly relevant to push activities, compared to PPR.

\subsubsection{On-line Evaluation}

In this section, we perform A-PARW-I and PPR algorithms in user selection for two different on-line push activities through Huawei's push service. Two lists of users are obtained by the two algorithms respectively, and the activity messages are push to these users \footnote{Duplicated users in the two lists are only pushed the message once.}. Interested users may click the message to see the details of the apps or perform further actions.

%The push service is an essential component of the real-time, always-on experience of Huawei phones. It offers efficient and reliable way of sending information to users. It allows users to be alerted of new and current information, e.g., updates of software or apps, or advertisements. In the typical scenario of on-line push activity, a pre-defined number of users are selected and then the activity messages are pushed to these users. If the push service is enabled on the user¡¯s phone as well as the phone is connected to the internet, then the messages can be delivered to the user¡¯s phone and the title of the message will be shown as an alert on the notification center. Interested users may click the message to see the details of the apps or perform further actions. User selection is a critical task in on-line advertisement, since too many unrelated ads must disturb users and degrade the user¡¯s experience of the phones.

\begin{figure}
\subfloat[CTR comparison]{
\label{online-click-ratio}
\includegraphics[width=.48\textwidth]{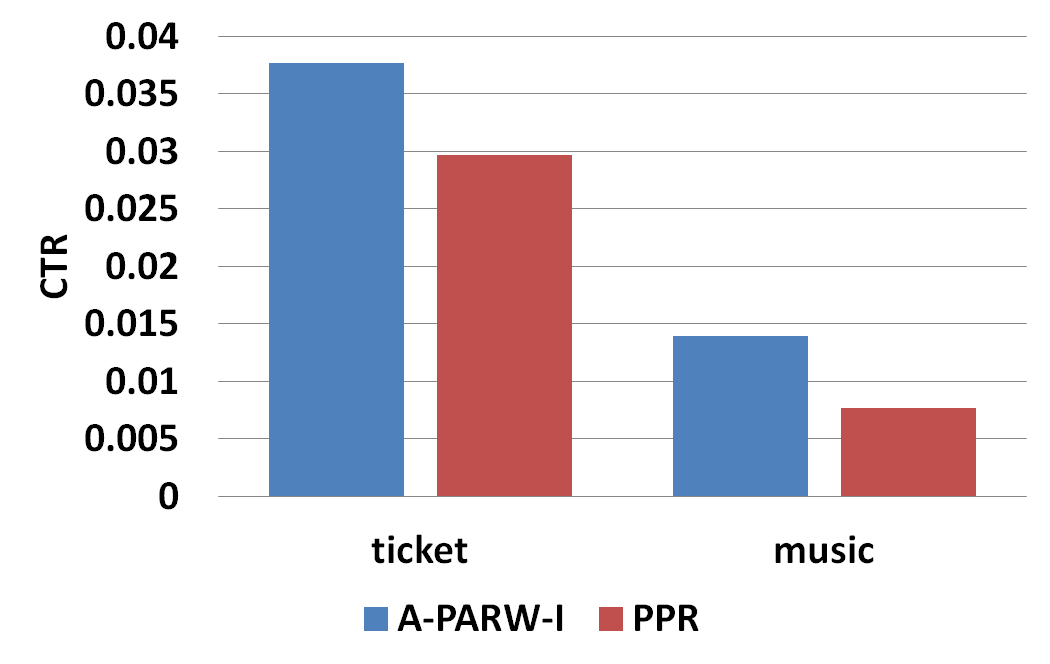}}
\subfloat[DTR comparison]{
\label{online-down-ratio}
\includegraphics[width=.48\textwidth]{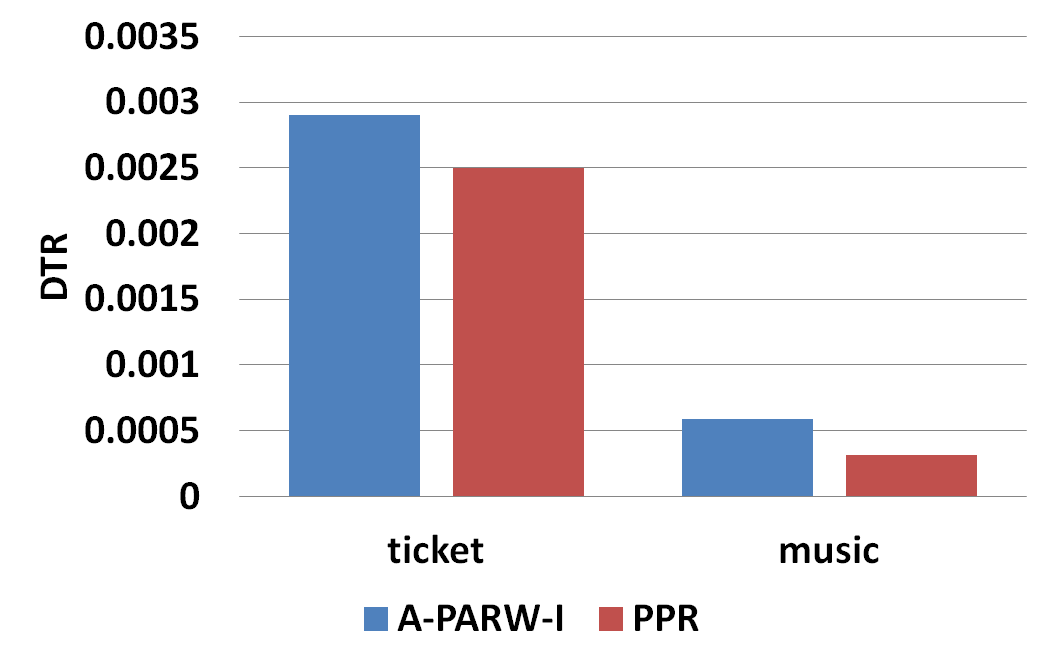}}
\caption{Online performance}
\label{fig:online}
\end{figure}

We calculated $CTR$ (click-through rate) and $DTR$ (download-through rate) from the log data of user feedbacks. For comparison, we define $CTR+$ and $DTR+$ as the improvement of A-PARW-I over PPR. The detail improvement values are presented in Table~\ref{tb:on-line improv}. More specifically, in Figure \ref{online-click-ratio} and \ref{online-down-ratio}, the $CTR$ and $DTR$ of A-PARW-I and PPR of the two on-line push service are shown. As we can see, the performance of A-PARW-I is significantly higher than PPR for both click and download in the two online push activities.

\begin{table}[h]
\centering
\caption{Online improvement of A-PARW-I over PPR }
\begin{tabular}{p{2cm}|p{2cm}|p{2cm}} \hline
        & ticket & music \\ \hline
$CTR+$   & 27\% & 82\%\\
$DTR+$    & 16\% & 85\%\\
\hline\end{tabular}
\label{tb:on-line improv}
\end{table}

\subsubsection{Property of PARW}\label{sec:expe:property}

In this subsection, we analyse the property of A-PARW-I and PPR from the view point of push service. The vital distinction of the two different A-PARW-based algorithms is that A-PARW-I pay more attention to graph community information and PPR over considers the degree of nodes in the graph. To verify this observation, we study the tendency of CTR/DTR and degree of the user-nodes selected by the two algorithms, sorting by the scores.
\begin{figure}
\centering
\subfloat[CTR tendency of activity ticket]{
\label{online-click-ticket}
\includegraphics[width=.40\textwidth]{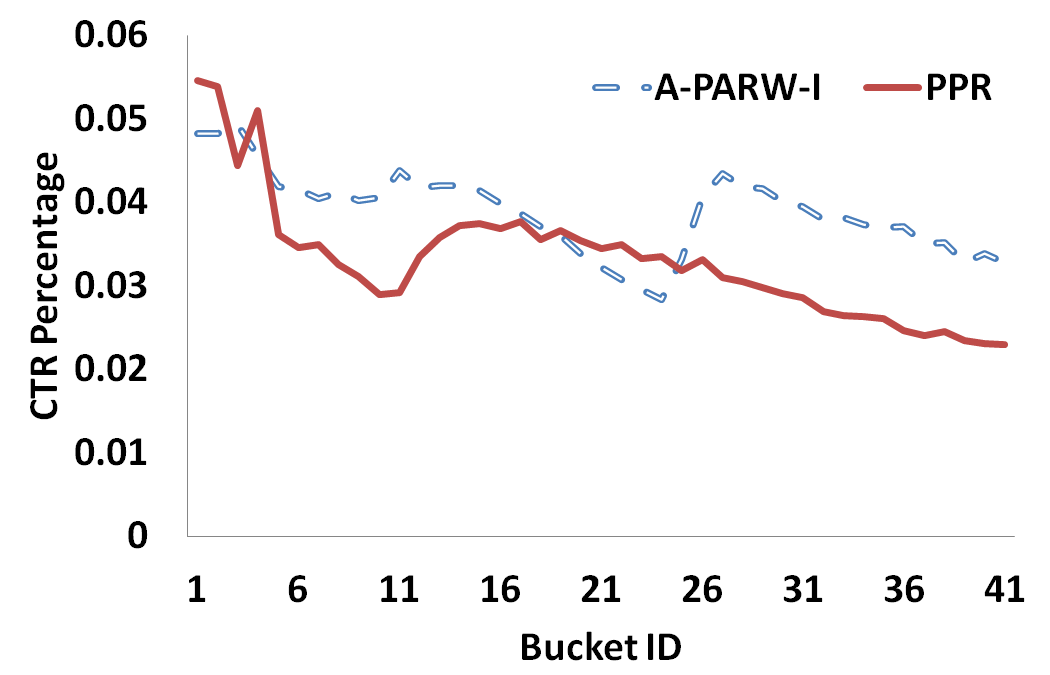}}
\subfloat[CTR tendency of activity music]{
\label{online-click-music}
\includegraphics[width=.40\textwidth]{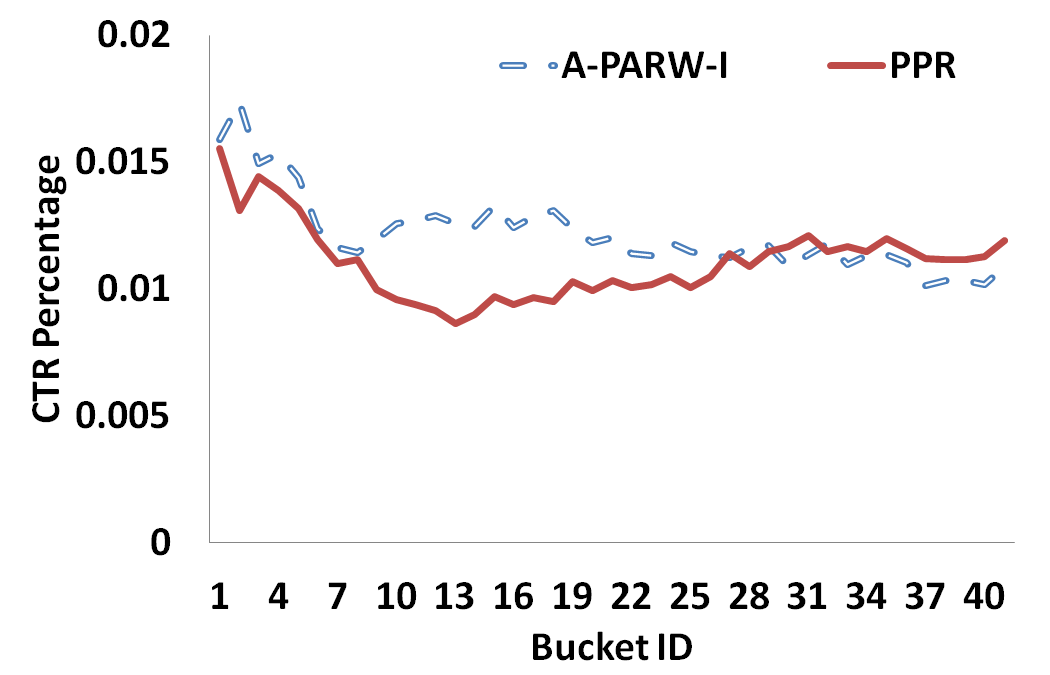}}\\
\subfloat[DTR tendency of activity ticket]{
\label{online-download-ticket}
\includegraphics[width=.40\textwidth]{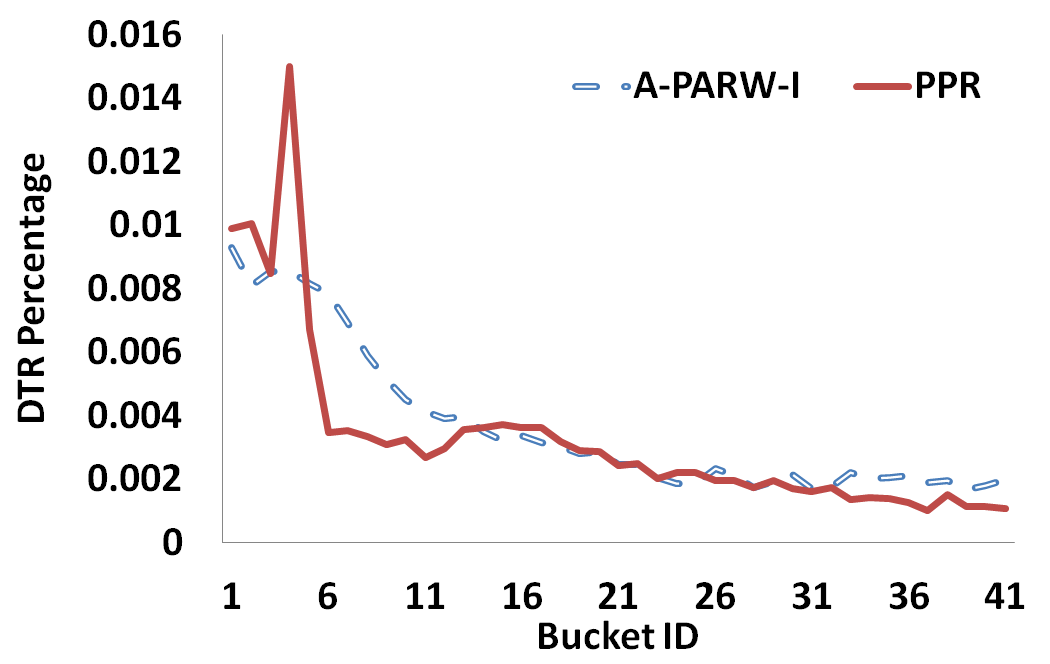}}
\subfloat[DTR tendency of activity music]{
\label{online-download-music}
\includegraphics[width=.40\textwidth]{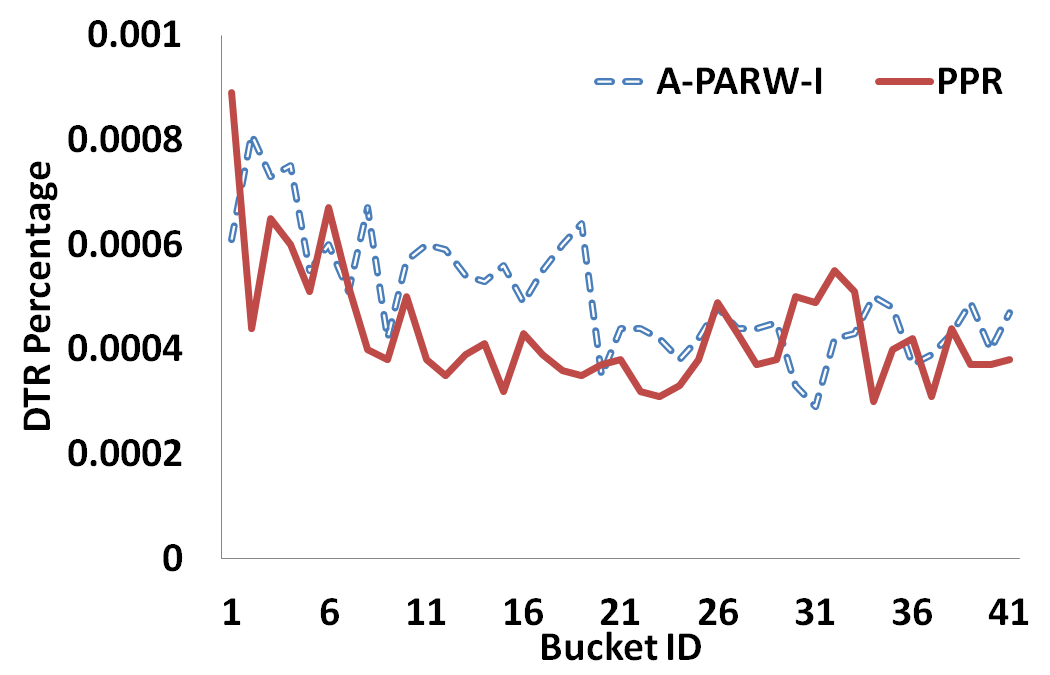}}
\caption{Tendency of $CTR$/$DTR$}
\label{fig:online-tendency}
\end{figure}

\begin{itemize}
\item \textbf{Tendency of CTR/DTR in sorted-steady-distribution}
Figure~\ref{fig:online-tendency} shows the tendency of $CTR$ and $DTR$ of two on-line push activities. More specifically, Figure~\ref{online-click-ticket}, Figure~\ref{online-click-music}, Figure~\ref{online-download-ticket} and Figure~\ref{online-download-music} represent tendency of activity ticket's $CTR$, activity music's $CTR$, ticket's $DTR$ and music's $DTR$. In the figures, x-axis is the bucket identifier (each bucket includes 100,000 users) and y-axis as the $CTR$ (respectively $DTR$) of the buckets. As we see, both PPR and A-PARW-I curves have high $CTR$ and $DTR$ value at beginning. However, PPR curve drops more dramatically than A-PARW-I curve; moreover,  PPR's $CTR$ and $DTR$ sometimes increase at the tail of the curve. Compared to PPR, A-PARW-I is more stable and steady, which means that A-PARW-I selects the users who are more relevant to the push activity and the ranking of the users are more reliable.

\item \textbf{Change of degree in sorted-steady-distribution} Figure~\ref{degree-ticket} and Figure~\ref{degree-music} present the degree's tendency of A-PARW-I and PPR in the activity of ticket and music respectively, where x-axis is the bucket identifier (each bucket includes 100,000 users) and y-axis is total degree of the users in the buckets. As we can see, the tendency of both red and blue curves are similar acorss the two figures. The PPR curve is higher than the A-PARW-I curve at first. However, PPR curve drops rapidly and A-PARW-I curve is more smooth and steady. It can be concluded from this two figures that PPR is prefers high-degree nodes while A-PARW-I considers global community structure.

\begin{figure}
\subfloat[degree tendency of activity ticket]{
\label{degree-ticket}
\includegraphics[width=.48\textwidth]{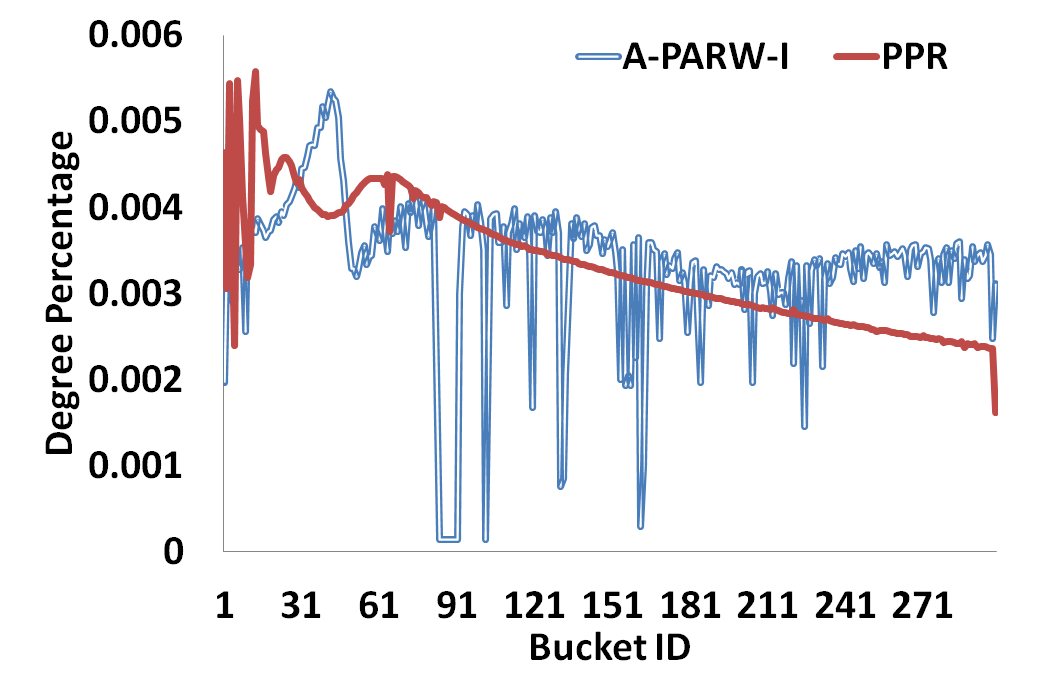}}
\subfloat[degree tendency of activity music]{
\label{degree-music}
\includegraphics[width=.48\textwidth]{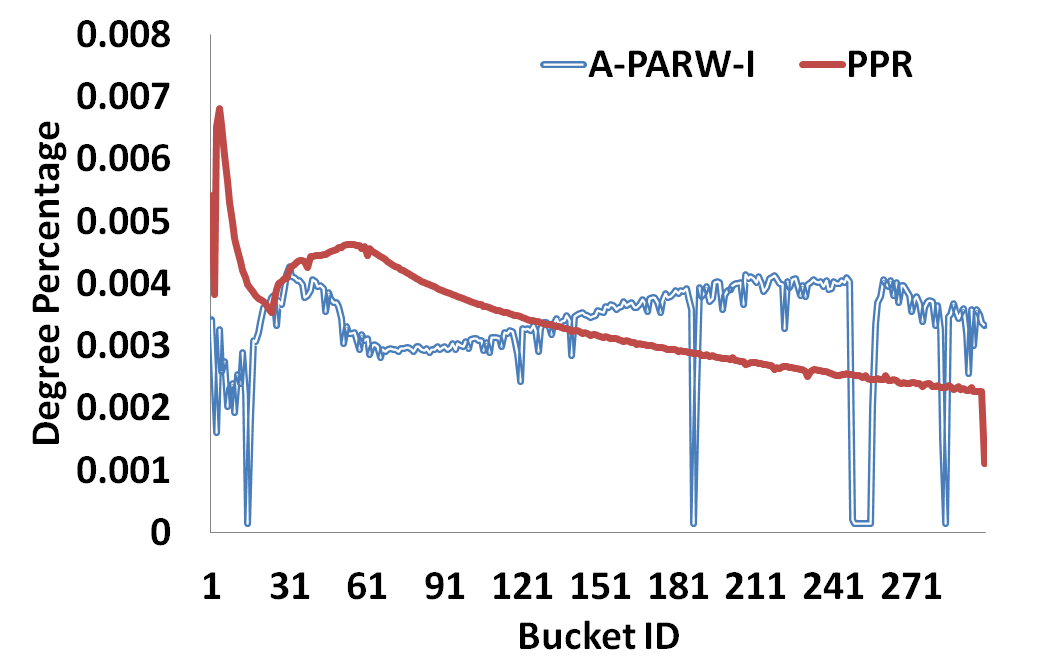}}
\caption{Tendency of users' degree}
\label{fig:change of degree}
\end{figure}

\end{itemize}

\section{Related Work}\label{sec:relatedwork}
%\subsection{pull-mode vs push-mode}\label{sec:rela:pullvspush}

%\textbf{Pull-mode vs push-mode}
In pull-mode recommendation, such as \cite{cheng2016wide,covington2016deep}, user could not receive latest updating before he/she enters Application Market. Compared to pull-mode recommendation, push-mode recommendation pushes specific messages to users according to its characteristics and does not require users to enter the system. The push recommendation with accurate user group selection is able to rebuild or strengthen the connection between Application Market and users.

The crucial part of push-mode is target user group discovery, which can be solved by \emph{rule-based} \cite{han2011data}, \emph{CF-based} \cite{su2009survey} and \emph{graph-based} \cite{he2015mining,li2008learning} approaches. However, the rule-based methods can not take advantage of collaborative information among users, because it needs a set of rules that is defined so that the accuracy rate is low and not flexible enough. The CF-based methods tend to be ineffective in real-world scenarios because it requires a great deal interactions of users and tags, which is very problematic due to the sparsity of data.

%\subsection{graph mining technique}\label{sec:rela:graph}
%\textbf{Graph mining technique}
In graph-based approach, PageRank \cite{page1999pagerank} is a well known link analysis algorithm used by Google to rank websites according to their importance. There are many variants of PageRank, such as TrustRank \cite{gyongyi2004combating} which is able to find non-spam pages, sensitive PageRank \cite{haveliwala2002topic} which mines personalized page and WTF \cite{gupta2013wtf} which is used to recommend friends in Twitter. However, PageRank based approach is biased to high-degree vertices. The authors of \cite{wu2012learning} propose a unified framework of graph mining which include a new algorithm PARW-I and personalized PageRank. PARW-I can capture the community structure to overcome the weakness of PageRank. We design a approximate version of PARW, namely A-PARW, in our system.
%shows a friends recommendation framework that based on random walk and SALSA, which is in the same family of random-walk algorithms as PageRank \cite{}.

%\input{Discussion.tex}
\section{Conclusions}\label{sec:conclu}
In this paper. we introduced Huawei Push Service Platform (shortly, PSP), a platform which performs push-mode recommendation by selecting target user group for a given push message, including potential users mining, online pushing, feedback caching and evaluation. In addition, we proposed a practical approximate version of PARW (namely A-PARW) for potential users mining. We presented a detail analysis between different modes of A-PARW algorithm based on our User-APP graph both theoretically and empirically. A-PARW captures community structure from User-APP graph and is not biased to high degree vertices, when $\varLambda=\alpha \cdot I$. We demonstrated that A-PARW-I is able to target the most relevant potential users and improves the performance of push service. As a live system, PSP supports the push-mode recommendation in Hispace Store.

%%%%%%%%%%%%%%%%%%%%%%%%%%%%%%%%%%%%%%%%%%%%%%%%%%%%%%%%%%%%%%%%%%%%%%%%%%%%%%%
\bibliographystyle{splncs03}
\bibliography{paper}

%All links were last followed on October 5, 2014.
%%%%%%%%%%%%%%%%%%%%%%%%%%%%%%%%%%%%%%%%%%%%%%%%%%%%%%%%%%%%%%%%%%%%%%%%%%%%%%%

\end{document}